\documentclass[aps,prd,reprint,twocolumn,10pt,eqsecnum,showpacs,nofootinbib,amsfonts,amssymb,amsmath,longbibliography,superscriptaddress,floatfix]{revtex4-1}
\usepackage[colorlinks=true,allcolors=blue]{hyperref}
\usepackage{bm}
\usepackage{color}
\usepackage{mathtools}
\usepackage{mathrsfs}
\usepackage{graphicx}

\DeclareSymbolFont{bbold}{U}{bbold}{m}{n}
\DeclareSymbolFontAlphabet{\mathbbold}{bbold}

\newcommand{\ud}{\mathrm{d}}
\newcommand{\uD}{\mathrm{D}}
\newcommand{\pb}[1]{\,\mbox{}_{#1}}
\newcommand{\vkappa}[1]{\smash{\overset{_\varkappa}{#1}}}
\newcommand{\half}[1]{\smash{\overset{_{\scalebox{0.4}{1/2}}}{#1}}}

\newcommand{\smallunderset}[2]{\smash{\underset{^{#1}}{#2}}}
\newcommand{\scri}{\mathscr{I}}

\usepackage[normalem]{ulem}

\begin{document}

\author{\'Eanna \'E.\ Flanagan}
\email{eef3@cornell.edu}
\affiliation{Department of Physics, Cornell University, Ithaca, New York, 14853, USA}
\author{Alexander M.\ Grant}
\email{amg425@cornell.edu}
\affiliation{Department of Physics, Cornell University, Ithaca, New York, 14853, USA}
\author{Abraham I.\ Harte}
\email{abraham.harte@dcu.ie}
\affiliation{Centre for Astrophysics and Relativity, School of Mathematical Sciences, Dublin City University, Glasnevin, Dublin 9, Ireland}
\author{David A.\ Nichols}
\email{d.a.nichols@uva.nl}
\affiliation{Gravitation Astroparticle Physics Amsterdam (GRAPPA), University of Amsterdam, Science Park, P.O.~Box 94485, 1090 GL Amsterdam, The Netherlands}

\title{Persistent gravitational wave observables: general framework}

%
%
\newcount\hh
\newcount\mm
\mm=\time
\hh=\time
\divide\hh by 60
\divide\mm by 60
\multiply\mm by 60
\mm=-\mm
\advance\mm by \time
\def\hhmm{\number\hh:\ifnum\mm<10{}0\fi\number\mm}

\begin{abstract}
  The gravitational wave memory effect is characterized by the permanent relative displacement of a pair of initially comoving test particles that is caused by the passage of a burst of gravitational waves.
  Recent research on this effect has clarified the physical origin and the interpretation of this gravitational phenomenon in terms of conserved charges at null infinity and ``soft theorems.''
  In this paper, we describe a more general class of effects than the gravitational wave memory that are not necessarily associated with these charges and soft theorems, but that are, in principle, measurable.
  We shall refer to these effects as \emph{persistent gravitational wave observables}.
  These observables vanish in nonradiative regions of a spacetime, and their effects ``persist'' after a region of spacetime which is radiating.
  We give three examples of such persistent observables, as well as general techniques to calculate them.
  These examples, for simplicity, restrict the class of nonradiative regions to those which are exactly flat.
  Our first example is a generalization of geodesic deviation that allows for arbitrary acceleration.
  The second example is a holonomy observable, which is defined in terms of a closed loop.
  It contains the usual ``displacement'' gravitational wave memory; three previously identified, though less well known memory effects (the proper time, velocity, and rotation memories); and additional new observables.
  Finally, the third example we give is an explicit procedure by which an observer could measure a persistent effect using a spinning test particle.
  We briefly discuss the ability of gravitational wave detectors (such as LIGO and Virgo) to measure these observables.
\end{abstract}

\maketitle

\tableofcontents

\section{Introduction}

The gravitational wave memory effect has, historically, been described as an enduring displacement between two nearby observers that can arise after gravitational waves pass by their positions.
Zel'dovich and Polnarev~\cite{Zeldovich1974} first noticed the effect in a calculation in linearized gravity of the fly-by of two astrophysical compact objects.
It has also been shown that the memory effect occurs in nonlinear general relativity~\cite{Christodoulou1991}; in this context, there is an additional effect (known as nonlinear or sometimes null memory~\cite{Bieri2014}) arising from perturbations generated by the effective stress-energy of the gravitational waves~\cite{Wiseman1991, Thorne1992} (see also~\cite{Tolish2014a}).
Even earlier, Newman and Penrose~\cite{Newman1966} had found that, near null infinity, surfaces of constant retarded time can have their shear set to zero as one approaches either spacelike or future timelike infinity, but not both; this is now understood as an aspect of the gravitational wave memory effect.

In addition to the gravitational scattering of compact objects, enduring displacements have been shown to occur in other astrophysical contexts: for example, in neutrino emission and kicks during core-collapse supernovae (e.g.,~\cite{Turner1978b, Burrows1996}), emission of matter during certain gamma-ray bursts (e.g.,~\cite{Segalis2001, Sago2004}), and compact-binary mergers (e.g.,~\cite{Favata2009, Pollney2010, Favata2011}).
Braginsky and Grishchuk~\cite{Braginsky1985} and Braginsky and Thorne~\cite{Braginsky1987} described the types of experiments needed to detect gravitational wave bursts with memory.
Searches for these bursts have been carried out using pulsar timing arrays, which have provided constraints on their frequency of occurrence~\cite{Wang2015, Arzoumanian2015}.
It may also be possible to detect the gravitational wave memory with the LIGO and Virgo detectors, once the detectors reach their design sensitivities~\cite{Lasky2016}.

There have also been recent developments in understanding the memory effect from a more theoretical perspective.
In particular, Strominger and collaborators have described a relationship between the memory effect, the soft theorem of Weinberg \cite{Weinberg1965}, and the supertranslation relating two specific Bondi frames before and after a burst of waves with memory (see the recent review~\cite{Strominger2017}, and references therein).
Additional effects in electromagnetism~\cite{Bieri2013, Pasterski2015b} and Yang-Mills theories~\cite{Pate2017} display this same property of association with soft theorems and asymptotic symmetries.
These ``triangles'' of memory effects, soft theorems, and the space of symmetries at null infinity (or recently, at boundaries of spacetime in general, such as the event horizon of a black hole~\cite{Hawking2016, Strominger2017, Compere2018b, Chandrasekaran2018}) have now become a key feature in the discussion of memory effects in the literature.

In this series of papers, we will be considering generalizations of the memory effect, not motivated by these theoretical considerations, but by what is measurable.
We consider the general class of what we call \emph{persistent gravitational wave observables}: namely, quantities that are the results of measurements which a set of observers can perform during some time interval, and that vanish if no gravitational waves have passed by.
Special cases of these observables are \emph{memory observables}, which we define as persistent observables that are associated with symmetries at spacetime boundaries.
The idea of persistent observables can be readily applied to contexts other than spacetime boundaries: for example, exact gravitational plane waves (e.g.~\cite{Harte2015, Zhang2017a}).
One issue in making the notion of a persistent observable precise is that there is no universal definition of a ``nonradiative'' region of a spacetime in which these observables would vanish, although such a definition exists in particular contexts, such as near null infinity or in linearized gravity with a fixed background.
In each context there exists a precise definition of persistent observables, but a persistent observable in one context may not satisfy the requirement that it vanish for nonradiative regions in another context, since the definitions of nonradiative differ.

In this paper, we will be considering the context where the spacetime is composed of three regions: two flat regions, with a curved region sandwiched in between containing gravitational waves.
The persistent observables in this case are the results of measurements that would yield trivial results if this curved region were also flat; as such, they can be thought of as ``integrated measures of curvature.''
Applying them to the special case where the curved region contains gravitational waves is what makes them persistent gravitational wave observables.
Although our results are in general nonlinear, these observables measure moments (in time) of the Riemann tensor and its derivatives along an observer's worldline when the curvature is weak.

In a subsequent paper, we will also consider observables that are defined near null infinity, where there is also an unambiguous notion of what one means by radiation.
In this context, the persistent observables can be expanded in a series in $1/r$, where $r$ is the Bondi-Sachs radial coordinate~\cite{Bondi1962, Sachs1962a}.
For the previously known persistent observables discussed in Sec.~\ref{sec:traditional}, the memory observables (the displacement, both leading and subleading) scale as $1/r$, whereas the persistent observables not associated with symmetries and conserved quantities (the relative velocity, rotation, and proper time observables) go as $1/r^2$.
We will check in a subsequent paper if this is also true for the new persistent observables defined in this paper.

A brief summary of the structure of this paper is as follows: first, in Sec.~\ref{sec:memory_overview}, we describe the persistent gravitational wave observables that we will be considering in this paper and their relationship to previous observables in the literature (particularly those in~\cite{Flanagan2014, Flanagan2016}).
For convenience, a simplified version of our results is presented in the same section.
We also briefly discuss how gravitational wave detectors might measure these observables.
The derivations of our results, in terms of covariant bitensors, are given in Secs.~\ref{sec:memory_bitensor} and~\ref{sec:memory_comp}; the former provides a review of techniques used in computations with covariant bitensors, and the latter gives the calculations themselves.
Further discussion and our conclusions are in Sec.~\ref{sec:discussion}.

We use the conventions for the metric and curvature tensors given in Wald~\cite{Wald1984}, the conventions for taking the dual of arbitrary tensors from Penrose and Rindler~\cite{Penrose1987, Penrose1988} (reviewed in Appendix~\ref{sec:dual}), and the conventions for bitensors from Poisson's review article~\cite{Poisson2004}.
We will use lowercase Latin letters for abstract spacetime tensor indices and capital Latin indices for tensor indices on an arbitrary vector bundle.
Lowercase Greek letters will be used to label components with respect to a parallel transported basis.
For brevity, we are using a convention for bitensors where we use the same annotations for indices as are used on the points at which the indices apply (e.g., $a$, $b$ at the point $x$ and $a'$, $b'$ at the point $x'$).
If a bitensor is a scalar at some point, we make the dependence on that point explicit.
Finally, for brevity, we will occasionally  take powers of order symbols, writing (e.g.) $O(a, b)^3$ as shorthand for $O(a^3, a^2 b, a b^2, b^3)$.

\section{Persistent gravitational wave observables} \label{sec:memory_overview}

\begin{table*}[ht]
  \centering
  \caption{\label{tab:memories} A summary of the persistent observables discussed in this paper.
    We provide the original reference for the observable (if it was defined before this paper), the section of this paper in which the observable is defined, and the equation in which we give the value of the observable (in the weak curvature limit).
    As a brief summary of the characteristics of these observables, we also give the number of time integrals of the Riemann tensor which appear in these observables (in the weak curvature and plane-wave limits; see Sec.~\ref{sec:feasibility} for more details) and the known scaling near null infinity in both the linearized theory and in full general relativity.
    If the observable is known to be associated with a known symmetry near a spacetime boundary (and so is a memory observable), that is indicated in the last column.}
  \bgroup
  \def\arraystretch{1.25}
  \begin{tabular}{lccccccccc} \hline \hline
    & & & & Number of time & \hspace{0.8em} & & & \hspace{0.8em} & Associated \\
    & & Definition & Result & integrals of the & & \multicolumn{2}{c}{Scaling near $\scri$ (if known)} & & with a known \\\cline{7-8}
    Observable & Reference & (Sec.) & (Eq.) & Riemann tensor & & Linearized GR & Full GR & & symmetry \\\hline
    Displacement & \cite{Zeldovich1974} & \ref{sec:traditional} & \eqref{eqn:disp_mem} & 2 & & $1/r$ & $1/r$ & & Yes \\
    Relative velocity & \cite{Grishchuk1989} & \ref{sec:traditional} & \eqref{eqn:vel_mem} & 1 & & $1/r^2$ & $\cdots$ & & No \\
    Relative rotation & \cite{Flanagan2014} & \ref{sec:traditional} & \eqref{eqn:rot_mem} & 1 & & $1/r^2$ & $\cdots$ & & No \\
    Relative proper time & \cite{Strominger2014b} & \ref{sec:traditional} & \eqref{eqn:time_mem} & 1 & & $1/r^2$ & $\cdots$ & & No \\
    Subleading displacement~\footnote{Subleading displacement memory near null infinity includes the spin memory~\cite{Pasterski2015a} and center of mass memory~\cite{Nichols2018}.} & \cite{Pasterski2015a, Nichols2018} & \ref{sec:traditional} & \eqref{eqn:subdisp_mem} & 3 & & $1/r$ & $1/r$ & & Yes \\
    Curve deviation & $\cdots$ & \ref{sec:curve_deviation_intro} & \eqref{eqn:curve_dev_vars},~\eqref{eqn:curve_dev_weak} & 1--3 \footnote{\label{foottab} With acceleration, the number of time integrals is 4 and higher.} & & $\cdots$ & $\cdots$ & & No \\
    Holonomy & $\cdots$ & \ref{sec:holonomy} & \eqref{eqn:holonomy_lambda_weak},~\eqref{eqn:holonomy_omega_weak} & 1--3 \textsuperscript{\ref{foottab}} & & $\cdots$ & $\cdots$ & & no \\
    Spinning test particle & $\cdots$ & \ref{sec:spinning_observables} & \eqref{eqn:spin_obs_decomp},~\eqref{eqn:spin_weak} & 1--2 & & $\cdots$ & $\cdots$ & & No \\\hline \hline
  \end{tabular}
  \egroup
\end{table*}

In this section, we will review previous examples of persistent gravitational wave observables in the literature; we then define our three new observables in Secs.~\ref{sec:curve_deviation_intro},~\ref{sec:holonomy}, and~\ref{sec:spinning_observables}; and finally, we give a brief discussion as to the feasibility of their measurement in~\ref{sec:feasibility}.
As mentioned in the Introduction, in all cases we assume that the regions before and after the burst of gravitational waves are flat.
A summary of the different persistent observables that occur in this paper is given in Table~\ref{tab:memories}.

\subsection{Traditional persistent gravitational wave observables} \label{sec:traditional}

The archetypal persistent gravitational observable is what we will call the \emph{displacement memory observable}: a change in proper distance between two initially comoving, unforced, and nearby observers before and after a burst of gravitational waves.
Denote the curves that the two observers follow by $\gamma$ and $\bar{\gamma}$, and the initial and final proper times of one of the observers by $\tau_0$ and $\tau_1$.
At each of $\tau_0$ and $\tau_1$, if $\bar{\gamma}$ is close enough, there is a unique geodesic that intersects both $\gamma$ and $\bar{\gamma}$ and is orthogonal\footnote{\label{fn:correspondence} Note that orthogonality is necessary to ensure that the geodesics are unique.
  There are different definitions of these geodesics that intersect $\gamma$ and $\bar{\gamma}$, which are called \emph{correspondences} in~\cite{Vines2014a}; we will not be using this particular correspondence past Sec.~\ref{sec:gen_holonomy}, instead using one which we will define in Sec.~\ref{sec:curve_deviation_intro}.} to $\gamma$.
We set the affine parameter $\lambda$ along these unique geodesics such that they intersect $\gamma$ at $\lambda = 0$ and $\bar{\gamma}$ at $\lambda = 1$.
The initial and final separation vectors $\xi^a$ and $\xi^{a'}$ are then the tangent vectors to these unique geodesics at $\gamma(\tau_0)$ and $\gamma(\tau_1)$, respectively.

Given these definitions, the change in separation can then be found explicitly by solving the geodesic deviation equation.
For initially comoving observers, this change is given by

\begin{equation} \label{eqn:disp_mem}
  \begin{split}
    \Delta \xi^\mu = &-\int_{\tau_0}^{\tau_1} \ud \tau_2 \int_{\tau_0}^{\tau_2} \ud \tau_3 R^\mu{}_{\alpha\nu\beta} (\tau_3) \dot{\gamma}^\alpha \dot{\gamma}^\beta \xi^\nu \\
    &+ O(\boldsymbol{\xi}^2, \boldsymbol{R}^2),
  \end{split}
\end{equation}
where $\dot\gamma^\alpha$ is the tangent vector to the curve $\gamma$.
In this expression, and all those that follow which use Greek indices, we are taking components on a basis that has been parallel transported along the worldline.

There is another type of displacement memory observable, one which depends instead on the initial relative velocity $\dot{\xi}^\alpha$ of the two worldlines.
This is the final relative displacement of two observers with no initial relative displacement, but an initial relative velocity.
An explicit expression, derived in a manner similar to Eq.~\eqref{eqn:disp_mem}, is given by

\begin{equation} \label{eqn:subdisp_mem}
  \begin{split}
    \widetilde{\Delta \xi}^\mu = &-\int_{\tau_0}^{\tau_1} \ud \tau_2 \int_{\tau_0}^{\tau_2} \ud \tau_3 \int_{\tau_3}^{\tau_2} \ud \tau_4 R^\mu{}_{\alpha\nu\beta} (\tau_4) \dot{\gamma}^\alpha \dot{\gamma}^\beta \dot{\xi}^\nu \\
    &+ O(\boldsymbol{\xi}^2, \boldsymbol{R}^2).
  \end{split}
\end{equation}
This we will call the \emph{subleading displacement memory observable}.
It is called subleading because of the additional time integral in Eq.~\eqref{eqn:subdisp_mem} as opposed to Eq.~\eqref{eqn:disp_mem}.
In the frequency domain, this time integral corresponds to multiplication by frequency or energy, and thus the quantity is subleading in the expansion in energy that is used in the corresponding soft theorems~\cite{Cachazo2014}.
The parts of the gravitational waves that produce the subleading displacement memory also arise at a higher order in the post-Newtonian expansion than the parts that generate the leading memory~\cite{Nichols2017, Nichols2018}.
This observable has been studied exclusively at null infinity, where it has been understood in terms of its electric and magnetic parity components, which are known as \emph{center of mass memory}~\cite{Nichols2018} and \emph{spin memory}~\cite{Pasterski2015a}, respectively.
The total subleading displacement is a memory observable, since both the spin and the center of mass memories are known to be associated with asymptotic symmetries (the spin memory is known to be associated with a soft theorem as well~\cite{Pasterski2015a}).

Since the geodesic deviation equation has solutions where the initially comoving observers have 4-velocities that become different over time, it is natural to wonder whether there could be a \emph{persistent relative velocity observable} given by a difference in the relative 4-velocities before and after a burst of gravitational waves; this is sometimes also referred to as the velocity memory in the literature~\cite{Grishchuk1989, Tolish2014a, Harte2015}.
It takes the form

\begin{equation} \label{eqn:vel_mem}
  \begin{split}
    \Delta \dot{\xi}^\mu = \frac{\uD}{\ud \tau_1} \Delta \xi^\mu = &-\int_{\tau_0}^{\tau_1} \ud \tau_2 R^\mu{}_{\alpha\nu\beta} (\tau_2) \dot{\gamma}^\alpha \dot{\gamma}^\beta \xi^\nu \\
    &+ O(\boldsymbol{\xi}^2, \boldsymbol{R}^2).
  \end{split}
\end{equation}
This effect is unavoidable in going between two regions that are initially and finally flat, at least for nonlinear plane waves~\cite{Bondi1989} (for a more recent discussion, see~\cite{Zhang2017a, Zhang2017b, Zhang2018}).
It has been suggested, moreover, that this relative velocity is in principle measurable for bursts generated by astrophysical sources~\cite{Grishchuk1989}.
Note that in the case where the final relative velocity is nonzero, the displacement memory will no longer be independent of the final time $\tau_1$, even after the burst of gravitational waves has passed.

Similarly, the observers can parallel transport orthonormal tetrads along their respective worldlines, and the tetrads are related to each other by a linear transformation that is influenced by the burst of gravitational waves.
The four-dimensional matrix representing this linear transformation is in general a Lorentz transformation; thus, we will call the effect a \emph{persistent Lorentz transformation observable}.
This matrix can be written as

\begin{equation} \label{eqn:lorentz_memory}
  \Lambda^\mu{}_\nu = \delta^\mu{}_\nu + \Delta \Omega^\mu{}_\nu,
\end{equation}
where

\begin{equation} \label{eqn:rot_mem}
  \Delta \Omega^\mu{}_\nu = \int_{\tau_0}^{\tau_1} \ud \tau_2 R^\mu{}_{\nu\alpha\beta} (\tau_2) \xi^\alpha \dot{\gamma}^\beta + O(\boldsymbol{\xi}^2, \boldsymbol{R}^2).
\end{equation}
Note this effect includes the relative velocity observable in the form of a boost (when contracted into $\dot{\gamma}^\nu$), as well as a \emph{relative rotation observable}.
The relative rotation observable could be inferred, for example, from integrating the equation of differential frame dragging (or differential Fokker precession) \cite{Estabrook1964, Nichols2011} once in time, although as far as we can tell no one has taken this approach before.

Finally, one might wonder if nearby geodesic observers measure the same amount of proper time elapsed along their worldlines after a burst of gravitational waves passes by their locations.
This proper time difference will depend on how the observers make their correspondence, in the sense of Footnote~\ref{fn:correspondence}.
Recall that the separation vectors defined above are tangents to unique geodesics that intersect both $\gamma$ and $\bar{\gamma}$ and are orthogonal to $\gamma$.
Suppose that the two observers synchronize their clocks such that the first of these geodesics (defining the initial separation) passes through $\gamma(\tau_0)$ and $\bar{\gamma} (\tau_0)$.
Then, the second of these geodesics (defining the final separation) will pass through the points $\gamma(\tau_1)$ and $\bar{\gamma} (\tau_1 + \Delta \tau)$; the proper time difference is this quantity $\Delta \tau$.
Strominger and Zhiboedov considered this observable in~\cite{Strominger2014b}, which we will call the \emph{persistent relative proper time observable}.
Performing a similar calculation as that which yields Eq.~\eqref{eqn:disp_mem}, we find that

\begin{equation} \label{eqn:time_mem}
  \Delta \tau = \frac{1}{2} \int_{\tau_0}^{\tau_1} \ud \tau_2 R_{\alpha\beta\gamma\delta} (\tau_2) \xi^\alpha \dot{\gamma}^\beta \xi^\gamma \dot{\gamma}^\delta + O(\boldsymbol{\xi}^3, \boldsymbol{R}^2).
\end{equation}

Near null infinity, the displacement memory scales with the Bondi-Sachs radial coordinate $r$ as $1/r$ (see, e.g.,~\cite{Strominger2017}).
We are not aware of calculations of the scaling of the relative velocity, rotation, and proper time observables with $r$ near null infinity, but otherwise in full generality.
Specializing to linearized gravity, however, we can make use of the results of~\cite{Bieri2014} to argue that these three persistent observables scale as $1/r^2$ (i.e., the term in the expansion at $1/r$ vanishes).\footnote{For spacetimes that are not asymptotically flat in the usual sense, see~\cite{Compere2018} for an example where the relative velocity does not have this scaling.}
In a future paper, we will study in greater detail the scaling with $r$ of these persistent observables and the new observables defined below.

Note that the persistent relative proper time, relative velocity, and relative rotation observables have all been called ``memories'' previously in the literature.
Since they are not associated with symmetries at boundaries of spacetime, we will be referring to them simply as persistent observables.

The observables we have discussed so far are all defined in a context where there are two flat regions of spacetime separated in time by a region with curvature.
One can also consider situations where there are two flat regions that are spatially separated, as for example occurs when considering the effects of intervening curvature on the propagation of null rays from sources to observers in astronomical observations.
In this context, a number of nonlocal observables can be defined (related to lensing, frequency shifts, etc.) (see, e.g., \cite{Harte2015, Grasso2018, Bartelmann2010, Hobbs2010}) which bear some similarities to the observables discussed here.

\subsection{Generalized holonomy} \label{sec:gen_holonomy}

In~\cite{Flanagan2014}, a covariant observable was introduced that encodes the four persistent gravitational wave observables of Sec.~\ref{sec:traditional} (displacement, velocity, proper time, and rotation) in a single vector.
We now review this observable, which was called the \emph{generalized holonomy} in~\cite{Flanagan2014}.
The generalized holonomy is based on the solutions $\chi^a$ of an \emph{affine transport} law along a curve with tangent vector $k^a$, which are given by solving the following differential equation along this curve:

\begin{equation} \label{eqn:affine_transport}
  k^a \nabla_a \chi^b = -k^b.
\end{equation}
If one solves Eq.~\eqref{eqn:affine_transport} with a given initial $\chi^a$ at some point $x$, then the final $\chi^{a'}$ at some point $x'$ along the curve can be written as follows:

\begin{equation} \label{eqn:affine_solns}
  \chi^{a'} = g^{a'}{}_a \chi^a + \Delta \chi^{a'}.
\end{equation}
The homogeneous solution $g^{a'}{}_a \chi^a$ corresponds to parallel transport of the given initial vector $\chi^a$.
Here $g^{a'}{}_a$ denotes the \emph{parallel propagator}, which we define in more detail in Sec.~\ref{sec:bitensors}.
The inhomogeneous solution $\Delta \chi^{a'}$ generalizes the notion of a separation vector between two points in flat spacetime.
In a curved spacetime, $\Delta \chi^{a'}$ and $g^{a'}{}_a$ depend on the curve connecting the points $x$ and $x'$.

Consider now a closed curve composed of two initially comoving timelike geodesics $\gamma$ and $\bar{\gamma}$ and two spatial geodesics connecting $\gamma$ and $\bar{\gamma}$ at the initial and final points of $\gamma$ and $\bar\gamma$.
Furthermore, assume that these spatial geodesics are both orthogonal to $\gamma$ at their respective points of intersection with $\gamma$.
Solve Eq.~\eqref{eqn:affine_transport} around this curve by starting at the initial point of $\gamma$, evolving forwards along $\gamma$, then along the geodesic connecting the final points, then backwards along $\bar{\gamma}$, and then finally along the geodesic between the initial points (this is the same orientation as given in Fig.~\ref{fig:holonomy}, which is introduced in Sec.~\ref{sec:holonomy}).
The solution~\eqref{eqn:affine_solns} defines a mapping

\begin{equation}
  \chi^a \to \Lambda^a{}_b \chi^b + \Delta \chi^a,
\end{equation}
called the generalized holonomy in~\cite{Flanagan2014}, where $\Lambda^a{}_b$ is the usual holonomy around this curve [as in Eq.~\eqref{eqn:lorentz_memory}].
The quantities $\Delta \chi^a$ and $\Lambda^a{}_b$ are the generalized holonomy observables.
It was shown in~\cite{Flanagan2014} that this generalized holonomy encodes the displacement memory, in addition to the persistent relative velocity, rotation, and proper time observables.
Specifically, the homogeneous solution encodes exactly the persistent Lorentz transformation observable $\Delta \Omega^\mu{}_\nu$ (in components with respect to a parallel transported basis), while the inhomogeneous solution encodes the displacement memory $\Delta \xi^\mu$, the persistent Lorentz transformation observable, and the persistent relative proper time observable $\Delta \tau$.
Explicit expressions are given by~\footnote{Note that this result is not the same equation as that given in~\cite{Flanagan2014} [their Eq.~(3.20)], because we are using a different initial point and direction for traversing the loop.
  The two results are consistent.}

\begin{subequations}
  \begin{align}
    \Lambda^\mu{}_\nu = &\;\delta^\mu{}_\nu + \Delta \Omega^\mu{}_\nu, \\
    \Delta \chi^\mu = &-\Delta \Omega^\mu{}_\nu \left[\xi^\nu + \Delta \xi^\nu + (\tau_1 - \tau_0) \dot{\gamma}^\nu\right] + \Delta \tau \dot{\gamma}^\mu \nonumber \\
    &- \Delta \xi^{\mu}.
  \end{align}
\end{subequations}

\subsection{Curve deviation} \label{sec:curve_deviation_intro}

\begin{figure}
  \includegraphics{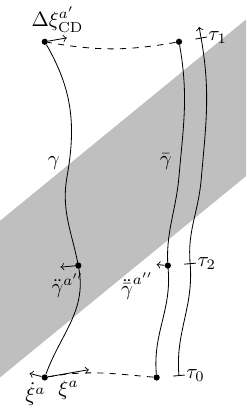}
  \caption{\label{fig:curve_dev} Two curves that have some initial separation $\xi^a$ and relative velocity $\dot{\xi}^a$, as well as accelerations $\ddot{\gamma}^{a''}$ and $\ddot{\bar \gamma}^{\bar a''}$.
  The curve deviation observable $\Delta \xi^{a'}_{\mathrm{CD}}$ is given by the difference between the measured final separation and the final separation that would be predicted if the gray region (containing gravitational waves) were also flat.}
\end{figure}

We now define our first new persistent observable---which we call \emph{curve deviation}---as a generalization of geodesic deviation (which forms the basis of the displacement memory observables).
Consider two timelike curves $\gamma$ and $\bar{\gamma}$ that pass through the region of gravitational waves, as depicted in Fig.~\ref{fig:curve_dev}.
Let $\tau_0$ denote a value of the affine parameter along $\gamma$ and $\bar \gamma$ before the gravitational waves have passed, and let $\tau_1$ be a value of this affine parameter after the passage of the waves.
The given variables in this problem are the initial separation $\xi^a$ and initial relative velocity $\dot{\xi}^a$ at $x \equiv \gamma(\tau_0)$, as well as the accelerations $\ddot{\gamma}^a$ and $\ddot{\bar \gamma}^{\bar a}$ at all values of $\tau$ along the curves (recall that we are using Latin indices to denote abstract indices).
At all later times $\tau$, we define the separation vector as tangent to the unique geodesic that connects $\gamma(\tau)$ and $\bar{\gamma} (\tau)$.
Note that this is not the same definition as in Sec.~\ref{sec:traditional}, where the separation was defined to always be orthogonal to $\dot{\gamma}^a$.
Since this definition connects two points on $\gamma$ and $\bar{\gamma}$ based upon having the same proper time, it is called the \emph{isochronous correspondence}, which is in contrast to the \emph{normal correspondence} which was used in Sec.~\ref{sec:traditional}.
For further discussion, see Sec.~\ref{sec:curve_deviation}.

The curve deviation observable $\Delta \xi^{a'}_{\textrm{CD}}$ at $x' \equiv \gamma(\tau_1)$ is the difference between the actual, measured separation, and the separation predicted from the observers' measured accelerations on a parallel transported basis, assuming that the region is flat.
Thus, this observable will vanish in flat spacetimes, even for arbitrarily accelerating curves.

Our observable has, in general, a nonlinear dependence on the initial separation and relative velocity, as well as the accelerations.
For simplicity, we parametrize the linear and quadratic dependence by the following quantities:

\begin{equation} \label{eqn:curve_dev_vars}
  \begin{split}
    \Delta \xi^{a'}_{\textrm{CD}} \equiv &\left[\Delta K^{a'}{}_b + L^{a'}{}_{bc} \xi^c + N^{a'}{}_{bc} \dot{\xi}^c + O(\boldsymbol{\xi}, \dot{\boldsymbol \xi})^2\right] \xi^b \\
    &+ \left[(\tau_1 - \tau_0) \Delta H^{a'}{}_b + M^{a'}{}_{bc} \dot{\xi}^c + O(\boldsymbol{\xi}, \dot{\boldsymbol \xi})^2\right] \dot{\xi}^b \\
    &+ \int_{\tau_0}^{\tau_1} \ud \tau_2 (\tau_1 - \tau_2) \Delta H^{a'}{}_{b''} \\
    &\hspace{7em}\times \Big\{g^{b''}{}_{\bar b''} \left[1 + O(\boldsymbol{\xi}, \dot{\boldsymbol \xi})^2\right] \ddot{\bar \gamma}^{\bar b''} \\
    &\hspace{8.5em}- \left[1 + O(\boldsymbol{\xi}, \dot{\boldsymbol \xi})^2\right] \ddot{\gamma}^{b''}\Big\} \\
    &+ O(\ddot{\boldsymbol \gamma}, \ddot{\bar{\boldsymbol \gamma}})^2.
  \end{split}
\end{equation}
Above we defined $x'' \equiv \gamma(\tau_2)$ and $g^{a''}{}_{\bar a''}$ as the bitensor which parallel transports vectors at $\bar{x}'' \equiv \bar{\gamma} (\tau_2)$ to $x''$.
The bitensors $\Delta K^{a'}{}_a$, $\Delta H^{a'}{}_a$, $L^{a'}{}_{bc}$, $M^{a'}{}_{bc}$, and $N^{a'}{}_{bc}$ vanish in flat spacetime and are determined by the curve $\gamma$ (and therefore depend, implicitly, on the acceleration $\ddot{\gamma}^a$).
Here, as mentioned in the Introduction, we are using $O(\boldsymbol{\xi}, \dot{\boldsymbol \xi})^2$ as shorthand for $O(\boldsymbol{\xi}^2, \boldsymbol{\xi} \cdot \dot{\boldsymbol \xi}, \dot{\boldsymbol \xi}^2)$.
A derivation of this result is given in Sec.~\ref{sec:curve_deviation}, and explicit expressions for all of the bitensors that this expression defines are given in Eq.~\eqref{eqn:curve_dev_results}.
The following are expressions for the bitensors in Eq.~\eqref{eqn:curve_dev_vars} that are valid to linear order in the Riemann tensor:

\begin{widetext}
\begin{subequations} \label{eqn:curve_dev_weak}
  \begin{align}
    \Delta K^\alpha{}_\beta &= -\int_{\tau_0}^{\tau_1} \ud \tau_2 \int_{\tau_0}^{\tau_2} \ud \tau_3 R^\alpha{}_{\gamma\beta\delta} (\tau_3) \dot{\gamma}^\gamma (\tau_3) \dot{\gamma}^\delta (\tau_3) + O(\boldsymbol{R}^2), \displaybreak[0] \\
    \Delta H^\alpha{}_\beta &= -\frac{1}{\tau_1 - \tau_0} \int_{\tau_0}^{\tau_1} \ud \tau_2 \int_{\tau_0}^{\tau_2} \ud \tau_3 \int_{\tau_3}^{\tau_2} \ud \tau_4 R^\alpha{}_{\gamma\beta\delta} (\tau_4) \dot{\gamma}^\gamma (\tau_4) \dot{\gamma}^\delta (\tau_4) + O(\boldsymbol{R}^2), \displaybreak[0] \\
    L^\alpha{}_{\beta\gamma} &= -\frac{1}{2} \int_{\tau_0}^{\tau_1} \mathrm{d} \tau_2 \int_{\tau_0}^{\tau_2} \mathrm{d} \tau_3 \{[\nabla_{(\epsilon} R^\alpha{}_{\gamma)\beta\delta}] (\tau_3) + [\nabla_{(\epsilon} R^\alpha{}_{\beta)\gamma\delta}] (\tau_3)\} \dot{\gamma}^\delta (\tau_3) \dot{\gamma}^\epsilon (\tau_3) + O(\boldsymbol{R}^2), \displaybreak[0] \\
    N^{\alpha}{}_{\beta\gamma} &= -\int_{\tau_0}^{\tau_1} \ud \tau_2 \int_{\tau_0}^{\tau_2} \ud \tau_3 \Bigg[\int_{\tau_3}^{\tau_2} \ud \tau_4 \left\{[\nabla_{(\epsilon} R^\alpha{}_{\gamma)\beta\delta}] (\tau_4) + [\nabla_{(\epsilon} R^\alpha{}_{\beta)\gamma\delta}] (\tau_4)\right\} \dot{\gamma}^\epsilon (\tau_4) \dot{\gamma}^\delta (\tau_4) + 2 R^\alpha{}_{\gamma\beta\delta} (\tau_3) \dot{\gamma}^\delta (\tau_3)\Bigg] \nonumber \\
    &\hspace{1em}+ O(\boldsymbol{R}^2), \displaybreak[0] \\
    M^{\alpha}{}_{\beta\gamma} &= \int_{\tau_0}^{\tau_1} \mathrm{d} \tau_2 \int_{\tau_0}^{\tau_2} \mathrm{d} \tau_3 \int_{\tau_3}^{\tau_2} \mathrm{d} \tau_4 \Bigg[\frac{1}{2} \int_{\tau_3}^{\tau_4} \mathrm{d} \tau_5 \{[\nabla_{(\epsilon} R^\alpha{}_{\gamma)\beta\delta}] (\tau_5) + [\nabla_{(\epsilon} R^\alpha{}_{\beta)\gamma\delta}] (\tau_5)\} \dot{\gamma}^\delta (\tau_5) \dot{\gamma}^\epsilon (\tau_5) \nonumber \\
    &\hspace{12.2em}- 2 R^\alpha{}_{(\gamma\beta)\delta} (\tau_4) \dot{\gamma}^\delta (\tau_4)\Bigg] \nonumber \\
    &- \frac{1}{2} \int_{\tau_0}^{\tau_1} \mathrm{d} \tau_2 \int_{\tau_2}^{\tau_1} \mathrm{d} \tau_3 \int_{\tau_0}^{\tau_3} \mathrm{d} \tau_4 \int_{\tau_4}^{\tau_3} \mathrm{d} \tau_5 \{[\nabla_{(\epsilon} R^\alpha{}_{\gamma)\beta\delta}] (\tau_5) + [\nabla_{(\epsilon} R^\alpha{}_{\beta)\gamma\delta}] (\tau_5)\} \dot{\gamma}^\delta (\tau_5) \dot{\gamma}^\epsilon (\tau_5) + O(\boldsymbol{R}^2).
  \end{align}
\end{subequations}
\end{widetext}
Note that the first of these expressions is very similar to Eq.~\eqref{eqn:disp_mem}, and the second is very similar to Eq.~\eqref{eqn:subdisp_mem}.
As in those equations, since we are integrating tensor fields, we note that we are considering the components along a parallel transported basis, which are denoted with Greek indices.
In these expressions, the parallel transported components of the 4-velocity are not constant functions of proper time, as $\gamma$ is not necessarily geodesic.

\subsection{Holonomies of linear and angular momentum} \label{sec:holonomy}

\begin{figure}
  \includegraphics{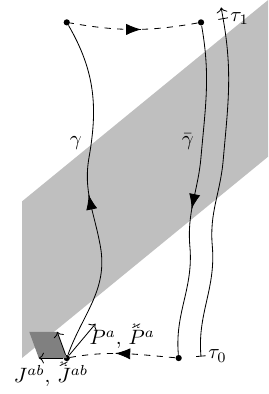}
  \caption{\label{fig:holonomy} A loop about which observers can compute a holonomy that measures the effect of a burst of gravitational waves.
    The quantities $P^a$ and $J^{ab}$ are transported around this loop using Eq.~\eqref{eqn:kappa_transport} (with some set of parameters $\varkappa$) in the directions shown, thereby yielding the observables $\vkappa{P}{}^a$ and $\vkappa{J}{}^{ab}$.}
\end{figure}

Our next observable is based on an extension of the affine transport law reviewed in Sec.~\ref{sec:gen_holonomy}, and it is motivated by the fact that this transport law also defines a means of relating linear and angular momenta at different points~\cite{Flanagan2014}.
Here, we mean either the linear and angular momentum of some extended body, or the linear and angular momentum of the spacetime itself.
There are a variety of prescriptions by which an observer could define linear and angular momentum, but what we will be considering in this paper is how the observer would sensibly transport these quantities from point to point.

As a motivating example, consider a freely falling body in flat spacetime.
Stated in terms of affine transport, an observer at some point $x$ would measure the total linear and angular momentum of the body (about her location) to be

\begin{equation} \label{eqn:J_decomposition}
  P^a \equiv g^a{}_{a'} P^{a'}, \quad J^{ab} \equiv g^a{}_{a'} g^b{}_{b'} S^{a'b'} + 2 \Delta \chi^{[a} P^{b]},
\end{equation}
respectively, where $S^{a'b'}$ and $P^{a'}$ are the intrinsic angular momentum and linear momentum of the body, and $\Delta \chi^a$ and $g^a{}_{a'}$ are the inhomogeneous and homogeneous solutions to Eq.~\eqref{eqn:affine_transport} given in Eq.~\eqref{eqn:affine_solns}.
The curve along which Eq.~\eqref{eqn:affine_transport} is being solved is the geodesic orthogonal to the observer's worldline that goes from a point $x'$ on the body's center of mass worldline to $x$.
The first term in this expression is the intrinsic angular momentum, and the second term is the orbital angular momentum, as it depends on the separation of the observer relative to the body.
Parallel propagators appear in this expression to reflect the fact that $P^{a'}$ and $S^{a'b'}$, as tensor fields, are only defined on the worldline of the freely falling spinning body.

A crucial feature of this example is that (when the intrinsic linear and angular momentum of the body are conserved) the constructed $P^a$ and $J^{ab}$ obey the following coupled differential equations along an arbitrary curve with tangent vector $k^a$:

\begin{subequations} \label{eqn:affine_transport_PJ}
  \begin{align}
    k^a \nabla_a P^b &= 0, \\
    k^a \nabla_a J^{bc} &= 2 P^{[b} k^{c]}.
  \end{align}
\end{subequations}
These differential equations capture the ``origin dependence'' of angular momentum in special relativity: that is, as we shift the origin (along some curve with tangent $k^a$) about which we are measuring the angular momentum, Eq.~\eqref{eqn:affine_transport_PJ} tells us how the linear and angular momentum change.
This origin dependence is a crucial feature of angular momentum, and so one could consider Eq.~\eqref{eqn:affine_transport_PJ} as the \emph{definition} of how linear and angular momentum should be transported in curved spacetimes (as was done in~\cite{Flanagan2014}), irrespective of the particular example used to derive this equation.
Unlike in flat spacetime, this transport depends on the curve that is used between the two points.

Solving Eq.~\eqref{eqn:affine_transport_PJ} around a closed loop gives a map from the space of $P^a$ and $J^{ab}$ to itself, which can be written in terms of the quantities $\Delta \chi^a$ and $\Lambda^a{}_b$ from the generalized holonomy.
In this paper, as in~\cite{Flanagan2016}, we will describe a generalization of this procedure for transporting linear and angular momentum; as the space of $P^a$ and $J^{ab}$ is ten-dimensional, our observable is a $10 \times 10$ matrix.
As a generalization of~\cite{Flanagan2016}, instead of Eq.~\eqref{eqn:affine_transport_PJ}, we solve the following differential equations along our curve:

\begin{subequations} \label{eqn:kappa_transport}
  \begin{align}
    k^b \nabla_b P^a &= -\vkappa{K}{}^a{}_{bcd} k^b J^{cd}, \\
    k^c \nabla_c J^{ab} &= 2 P^{[a} k^{b]}.
  \end{align}
\end{subequations}
Here $\varkappa = (\varkappa_1, \varkappa_2, \varkappa_3, \varkappa_4)$ is a collection of constant parameters, and the tensor $\vkappa{K}{}^{ab}{}_{cd}$ is defined by

\begin{equation} \label{eqn:K_def}
  \begin{split}
    \vkappa{K}{}^{ab}{}_{cd} = \varkappa_1 R^{ab}{}_{cd} &+ \varkappa_2 \delta^a{}_{[c} R^b{}_{d]} + \varkappa_3 \delta^b{}_{[c} R^a{}_{d]} \\
    &+ \varkappa_4 R \delta^a{}_{[c} \delta^b{}_{d]}.
  \end{split}
\end{equation}
Two special cases of this transport law have been considered in previous work: $\varkappa = (0, 0, 0, 0)$, which was studied in~\cite{Flanagan2014}, and $\varkappa = (\kappa, 0, 0, 0)$, which was studied in~\cite{Flanagan2016}.

Our \emph{holonomy observable} will be given by solving Eq.~\eqref{eqn:kappa_transport} around a closed loop.
This gives us a curve-dependent observable (in the form of a matrix at a given point) describing how the final linear and angular momentum, which we denote by $\vkappa{P}{}^a$ and $\vkappa{J}{}^{ab}$, depend on the initial linear and angular momentum, which we denote by $P^a$ and $J^{ab}$.
We call this matrix the holonomy, and it can be decomposed into four components:

\begin{equation} \label{eqn:holonomy_vars}
  \begin{pmatrix}
    \vkappa{P}{}^a \\
    \vkappa{J}{}^{ab}
  \end{pmatrix} = \begin{pmatrix}
    \smallunderset{PP}{\vkappa \Lambda}{}^a{}_c & \smallunderset{PJ}{\vkappa \Lambda}{}^a{}_{cd} \vspace{0.5em} \\
    \smallunderset{JP}{\vkappa \Lambda}{}^{ab}{}_c & \smallunderset{JJ}{\vkappa \Lambda}{}^{ab}{}_{cd}
  \end{pmatrix} \begin{pmatrix}
    P^c \\
    J^{cd}
  \end{pmatrix}.
\end{equation}
This observable depends on the \emph{curve} used in its definition.
For example, in~\cite{Flanagan2016}, this holonomy was computed for infinitesimal square loops.
In this paper, the holonomy will be computed for the case of a narrow loop, as in Fig.~\ref{fig:holonomy}, where two of the edges are much shorter than the curves $\gamma$ and $\bar{\gamma}$.

The dependence on the curve is apparent in two ways.
First, the holonomy depends on the separation vector $\xi^a$ throughout the curved region, so it depends on the initial separation, relative velocity, and the accelerations of the curves.
This will be made more concrete in Eq.~\eqref{eqn:holonomy_lambda_weak} (for the weak curvature case), Eq.~\eqref{eqn:holo_narrow_soln} (in general), and in the discussion in Sec.~\ref{sec:narrow}.
Second, even given the curves $\gamma$ and ${\bar \gamma}$, the holonomy depends on the choices of initial and final points on $\gamma$ that define the closed curve.
However, it must be noted that because we have assumed that the regions are flat before and after the burst of gravitational waves, the dependence on the start or end of the loop is in some sense ``trivial'': namely, it is only related to the usual origin dependence of angular momentum in special relativity.

We now consider particular values of $\varkappa$.
As mentioned above, the holonomy for $\varkappa = (0, 0, 0, 0)$ can be written in terms of $\Lambda^a{}_b$ and $\Delta \chi^a$:

\begin{subequations}
  \begin{align}
    \smallunderset{PP}{\mathring \Lambda}{}^a{}_c &= \Lambda^a{}_c, &\quad \smallunderset{PJ}{\mathring \Lambda}{}^a{}_{cd} &= 0, \\
    \smallunderset{JP}{\mathring \Lambda}{}^{ab}{}_c &= 2 \Delta \chi^{[a} \Lambda^{b]}{}_c, \quad &\smallunderset{JJ}{\mathring \Lambda}{}^{ab}{}_{cd} &= \Lambda^{[a}{}_c \Lambda^{b]}{}_d.
  \end{align}
\end{subequations}
Thus, the value of this holonomy has already been effectively computed in~\cite{Flanagan2014}, and (as noted in Sec.~\ref{sec:gen_holonomy}) the components of this holonomy describe the usual displacement memory observable, as well as the relative velocity and rotation observables.
We will be computing this observable again, using a different framework, in Sec.~\ref{sec:affine}.

In addition to the case of $\varkappa = (0, 0, 0, 0)$, which we will refer to as \emph{affine transport}, we also consider the case $\varkappa = (1/2, 0, 0, 0)$, which we will refer to as \emph{dual Killing transport}.
The holonomy of dual Killing transport describes how the space of symmetries changes because of the burst of gravitational waves.
This is due to the relationship between the transport law in Eq.~\eqref{eqn:kappa_transport} with $\varkappa = (1/2, 0, 0, 0)$ and the Killing transport equations which determine how Killing vector fields can be determined from initial data at a point.
This can be seen as follows: for any Killing vector field $\xi^a$, define $\omega_{ab} \equiv \nabla_{[a} \xi_{b]}$.
The Killing transport equations can then be written as

\begin{subequations} \label{eqn:killing_transport}
  \begin{align}
    \nabla_a \xi_b &= \omega_{ab}, \\
    \nabla_a \omega_{bc} &= R^d{}_{abc} \xi_d.
  \end{align}
\end{subequations}
This implies that, for $P^a$ and $J^{ab}$ transported along a curve by dual Killing transport,

\begin{equation} \label{eqn:Q_def}
  Q = P^a \xi_a + \frac{1}{2} J^{ab} \omega_{ab}
\end{equation}
is a constant (see, e.g.,~\cite{Harte2008}).

This association between linear and angular momentum and Killing vector fields allows us to think of the holonomy in a slightly different way.
The Killing transport equations~\eqref{eqn:killing_transport} can be used to transport an element of the space of symmetries in the flat region before the burst along the curve $\gamma$ by contracting the free index $a$ with $\dot{\gamma}^a$ (to yield ordinary differential equations along $\gamma$).
This gives an element of the space of symmetries in the region after the burst, thus providing a linear map between the spaces of symmetries before and after the burst.
This map is independent of the initial and final points along $\gamma$ between which this transport is carried out, so long as the points are within the flat regions.
There is a corresponding map along $\bar{\gamma}$ that maps the space of symmetries after the burst to the space of symmetries before.
Composing these maps yields a holonomy that is, in a sense, ``dual'' to the holonomy discussed in this section.
This holonomy maps from the space of initial Killing vector fields to the space of final Killing vector fields and is independent of the choices of initial and final points on the curves $\gamma$ and $\bar{\gamma}$.
However, the components of this map on a basis that is determined by the initial point (that is, components of angular momentum about the initial point on $\gamma$) or the final point depend on those choices.

Finally, there is a third relevant value of $\varkappa$, namely $\varkappa = (-1/4, 1/2, 0, 0)$.
This is the value of $\varkappa$ that is most interesting near null infinity, as discussed in~\cite{Flanagan2016}, since it turns out that the holonomy with this value of $\varkappa$ is trivial in an asymptotic sense in stationary regions.
We will be discussing this in more detail in a future paper.

In summary, the holonomy observable discussed in this section provides a persistent gravitational wave observable which we will compute in this and subsequent papers.
Our final result for the value of the matrix in Eq.~\eqref{eqn:holonomy_vars} is in Eq.~\eqref{eqn:0_holonomy_final} for the case of affine transport and~\eqref{eqn:half_Lambda} for the case of dual Killing transport.
For a general set of parameters $\varkappa$, our results are given in Eqs.~\eqref{eqn:kappa_holonomy} and~\eqref{eqn:kappa_hol_Omegas}.
For convenience, we list our results in the weak curvature limit below for general $\varkappa$:

\begin{widetext}
\begin{equation} \label{eqn:holonomy_lambda_weak}
  \begin{split}
    \begin{pmatrix}
      \smallunderset{PP}{\vkappa{\Lambda}}{}^\alpha{}_\mu & \smallunderset{PJ}{\vkappa{\Lambda}}{}^\alpha{}_{\mu\nu} \vspace{0.5em} \\
      \smallunderset{JP}{\vkappa{\Lambda}}{}^{\alpha\beta}{}_\mu & \smallunderset{JJ}{\vkappa{\Lambda}}{}^{\alpha\beta}{}_{\mu\nu}
    \end{pmatrix} = &\begin{pmatrix}
      \delta^\alpha{}_\mu & 0 \\
      0 & \delta^{[\alpha}{}_\mu \delta^{\beta]}{}_\nu
    \end{pmatrix} \\
    &+ \int_{\tau_0}^{\tau_1} \ud \tau_2 \Bigg\{\begin{pmatrix}
      \smallunderset{PP}{\Omega}{}^\alpha{}_{\mu\kappa} (\tau_1, \tau_2) & \smallunderset{PJ}{\Omega}{}^\alpha{}_{\mu\nu\kappa} (\tau_2) \vspace{0.25em} \\
       \smallunderset{JP}{\Omega}{}^{\alpha\beta}{}_{\mu\kappa} (\tau_1, \tau_2) & \smallunderset{JJ}{\Omega}{}^{\alpha\beta}{}_{\mu\nu\kappa} (\tau_1, \tau_2)
     \end{pmatrix} \xi^\kappa + \int_{\tau_2}^{\tau_1} \ud \tau_3 \begin{pmatrix}
       \smallunderset{PP}{\Omega}{}^\alpha{}_{\mu\kappa} (\tau_1, \tau_3) & \smallunderset{PJ}{\Omega}{}^\alpha{}_{\mu\nu\kappa} (\tau_3) \vspace{0.25em} \\
        \smallunderset{JP}{\Omega}{}^{\alpha\beta}{}_{\mu\kappa} (\tau_1, \tau_3) & \smallunderset{JJ}{\Omega}{}^{\alpha\beta}{}_{\mu\nu\kappa} (\tau_1, \tau_3)
     \end{pmatrix} \dot{\xi}^\kappa \\
                                                          &\hspace{5em}+ \left[\ddot{\bar \gamma}^\kappa (\tau_2) - \ddot{\gamma}^\kappa (\tau_2) \right] \int_{\tau_2}^{\tau_1} \ud \tau_3 \int_{\tau_3}^{\tau_1} \ud \tau_4 \begin{pmatrix}
                                                            \smallunderset{PP}{\Omega}{}^\alpha{}_{\mu\kappa} (\tau_1, \tau_4) & \smallunderset{PJ}{\Omega}{}^\alpha{}_{\mu\nu\kappa} (\tau_4) \vspace{0.25em} \\
                                                            \smallunderset{JP}{\Omega}{}^{\alpha\beta}{}_{\mu\kappa} (\tau_1, \tau_4) & \smallunderset{JJ}{\Omega}{}^{\alpha\beta}{}_{\mu\nu\kappa} (\tau_1, \tau_4)
                                                          \end{pmatrix}\Bigg\} \\
                                                                                      &+ O(\boldsymbol{\xi}, \dot{\boldsymbol{\xi}})^2 + O(\boldsymbol{R}, \boldsymbol{\nabla R})^2,
  \end{split}
\end{equation}
where

\begin{subequations} \label{eqn:holonomy_omega_weak}
  \begin{align}
    \smallunderset{PP}{\Omega}{}^\alpha{}_{\mu\kappa} (\tau_1, \tau_2) &= \left\{R^\alpha{}_{\mu\kappa\lambda} (\tau_2) + 4 \vkappa{K}{}^\alpha{}_{(\mu|\lambda|\kappa)} (\tau_2) + 2 \int_{\tau_2}^{\tau_1} \ud \tau_3 [\nabla_\kappa \vkappa{K}{}^\alpha{}_{\sigma\mu\lambda}] (\tau_3) \dot{\gamma}^\sigma (\tau_3)\right\} \dot{\gamma}^\lambda (\tau_2), \displaybreak[0] \\
    \smallunderset{PJ}{\Omega}{}^\alpha{}_{\mu\nu\kappa} (\tau_1) &= 2 [\nabla_{[\kappa} \vkappa{K}{}^\alpha{}_{\lambda]\mu\nu}] (\tau_1) \dot{\gamma}^\lambda (\tau_1), \displaybreak[0] \\
    \smallunderset{JP}{\Omega}{}^{\alpha\beta}{}_{\mu\kappa} (\tau_1, \tau_2) &= 8 \int_{\tau_2}^{\tau_1} \ud \tau_3 \mathbf{\Bigg(}\delta^{[\alpha}{}_\rho \dot{\gamma}^{\beta]} (\tau_2) \Bigg\{[\nabla_{[\kappa} \vkappa{K}{}^\rho{}_{\lambda]\mu\zeta}] (\tau_3) \int_{\tau_0}^{\tau_2} \ud \tau_4 \dot{\gamma}^\zeta (\tau_4) - \vkappa{K}{}^\rho{}_{(\mu|\lambda|\kappa)} (\tau_3) \nonumber \\
    &\hspace{11.5em}- \frac{1}{2} \int_{\tau_3}^{\tau_1} \ud \tau_4 [\nabla_\kappa \vkappa{K}{}^\rho{}_{\sigma\mu\lambda}] (\tau_4) \dot{\gamma}^\sigma (\tau_4)\Bigg\} \nonumber \\
    &\hspace{6em}+ \frac{1}{4} \dot{\gamma}^\sigma (\tau_2) \Big[\delta^{[\alpha}{}_\mu R^{\beta]}{}_{\sigma\kappa\lambda} (\tau_3) + 4 \delta^{[\alpha}{}_{[\kappa} \vkappa{K}{}^{\beta]}{}_{\lambda]\mu\sigma} (\tau_3)\Big]\mathbf{\Bigg)} \dot{\gamma}^\lambda (\tau_3), \displaybreak[0] \\
    \smallunderset{JJ}{\Omega}{}^{\alpha\beta}{}_{\mu\nu\kappa} (\tau_1, \tau_2) &= 2 \dot{\gamma}^{[\alpha} (\tau_2) \int_{\tau_2}^{\tau_1} \ud \tau_3 [\nabla_\kappa \vkappa{K}{}^{\beta]}{}_{\lambda\mu\nu}] (\tau_3) \dot{\gamma}^\lambda (\tau_3) + 2 \Big[\delta^{[\alpha}{}_{[\mu} R^{\beta]}{}_{\nu]\kappa\lambda} (\tau_2) + \delta^{[\alpha}{}_\kappa \vkappa{K}{}^{\beta]}{}_{\lambda\mu\nu} (\tau_2)\Big] \dot{\gamma}^\lambda (\tau_2).
  \end{align}
\end{subequations}
\end{widetext}

The explicit expressions for affine transport [$\varkappa = (0, 0, 0, 0)$] and dual Killing transport [$\varkappa = (1/2, 0, 0, 0)$] are apparent, and are given in more detail in Secs.~\ref{sec:affine} and~\ref{sec:dual_killing}.
These expressions are quite complicated for the case of general $\varkappa$, even in this weak curvature limit.
In the case of affine transport, we know that the results are far simpler, and that there are fewer independent components.
In Appendix~\ref{sec:algebraic}, we will consider a way of decomposing the holonomy into different parts, which may aid in understanding the meaning of the large number of components in the holonomy.
In the case of affine transport, this decomposition makes the reduction in the number of independent components manifest.

\subsection{Observables involving a spinning test particle} \label{sec:spinning_observables}

\begin{figure}
  \includegraphics{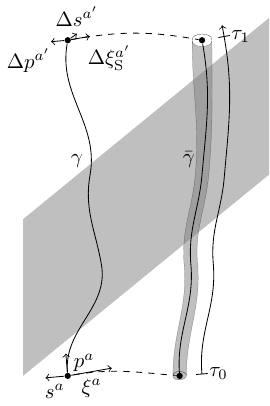}
  \caption{\label{fig:observable} An observer (left curve, $\gamma$) measuring properties of a spinning test particle (right curve, $\bar{\gamma}$).
    The test particle has some measured separation $\xi^a$, linear momentum $p^a$, and intrinsic spin $s^a$ before a burst of gravitational waves.
    After the burst, these quantities are all measured again and compared with their values before the burst, yielding $\Delta \xi^{a'}$, $\Delta p^{a'}$, and $\Delta s^{a'}$.}
\end{figure}

The holonomy given in the previous section is a powerful mathematical map for determining how a radiative region has affected how an observer keeps track of some angular momentum that she has measured.
However, it is an abstract quantity.
In particular, it depends on a closed \emph{curve} that intersects the curved region, and it could be difficult to measure in practice for some arbitrary curve.
The curve deviation observable given in Sec.~\ref{sec:curve_deviation_intro} is a more realistically observable quantity, but it also requires the observers to measure their acceleration at all times.
A more ideal observable would be one that would only require measurements before and after the burst of gravitational waves.
An example of such an observable is given by the following procedure.
An unaccelerated observer measures the linear momentum and intrinsic spin of a comoving test particle, in addition to its separation from the observer.
The observer and the test particle then travel along their own worldlines, and after the burst of gravitational waves, the separation, linear momentum, and intrinsic spin of the test particle are measured again.
The procedure is depicted in Fig.~\ref{fig:observable}, where we denote the worldlines of the observer and spinning test particle by $\gamma$ and $\bar{\gamma}$, respectively.
The differences between the initial and final separations, linear momenta, and spins per unit mass are the natural observables in this procedure.

We note that both the linear momentum and the intrinsic spin are tensors, and therefore, unless the initial and final separations are zero, we must specify a prescription for transporting these tensors away from the worldline of the spinning particle.
The convention which we use is that both are parallel transported along a curve connecting the two worldlines.
Since the regions before and after the burst are flat, this procedure is independent of the particular curves used.
Moreover, the measurements of linear momentum, intrinsic spin, and the separation are all tensors at specific points along the observer's worldline, so they must be parallel transported along the observer's worldline to some common point in order to be compared.

Note that this procedure is qualitatively similar to the holonomy for dual Killing transport, as the linear and angular momentum of the spinning test particle evolve along its worldline according to the Mathisson-Papapetrou equations~\cite{Mathisson1937, Papapetrou1951}:

\begin{subequations} \label{eqn:mp}
  \begin{align}
    \dot{\bar \gamma}^{\bar b} \nabla_{\bar b} p^{\bar a} &= -\frac{1}{2} R^{\bar a}{}_{\bar b \bar c \bar d} \dot{\bar \gamma}^{\bar b} j^{\bar c \bar d}, \label{eqn:mp_Pdot} \\
    \dot{\bar \gamma}^{\bar c} \nabla_{\bar c} j^{\bar a \bar b} &= 2 p^{[\bar a} \dot{\bar \gamma}^{\bar b]}. \label{eqn:mp_Jdot}
  \end{align}
\end{subequations}
Here $p^{\bar a}$ and $j^{\bar a \bar b}$ are the linear and angular momentum of the spinning particle when measured about $\bar{\gamma} (\tau)$.
These equations are precisely the transport law for $p^{\bar a}$ and $j^{\bar a \bar b}$ using dual Killing transport.
However, note that $p^{\bar a}$ and $j^{\bar a \bar b}$ are parallel transported along the geodesics connecting the two worldlines during the measurement process, and when the observer compares her initial and final measurements.
This observable, therefore, cannot be understood in terms of a holonomy, as different transport laws are used along different portions of the loop.

The second difference between this procedure and the holonomy is that the holonomy can be computed around an arbitrary loop, whereas in this procedure the worldlines $\gamma$ and $\bar \gamma$ are more constrained.
The curve $\bar{\gamma}$ here refers to a ``reference worldline'' for the spinning test particle.
This worldline is arbitrary, in a sense, and refers to the center of mass of the particle defined by certain \emph{spin-supplementary conditions} (for a review, see~\cite{Kyrian2007} and the references therein).
We will discuss our choice of spin-supplementary condition further in Sec.~\ref{sec:spin_comp}, as it is crucial for determining the acceleration of the spinning particle and therefore the exact shape of the loop.
Related to our choice of spin supplementary condition is our definition of the intrinsic spin per unit mass, which we discuss in the same section.

Explicitly, we denote our observables, the differences between initial and final separation, measured linear momenta, and measured intrinsic spins per unit mass of the test particle, by $\Delta \xi^{a'}_{\textrm S}$, $\Delta p^{a'}$, and $\Delta s^{a'}$, respectively.
These are functions of the initially measured 4-momentum $p^a$, initial intrinsic spin per unit mass $s^a$, and initial separation $\xi^a$.
We expand both in the separation and in the intrinsic spin, assuming that the size of the body, which is characterized by the spin per unit mass, is small as well.
That is, our approximation is that

\begin{equation}
  |\boldsymbol{\xi}| \gtrsim |\boldsymbol{s}| \gg m,
\end{equation}
where $m$ is the mass of the particle.
The assumption that the spin per unit mass is much larger than the mass is necessary in order to neglect the effects of self-force on the test particle.

To second order in separation, but first order in intrinsic spin, we can write

\begin{subequations} \label{eqn:spin_obs_decomp}
  \begin{align}
    \Delta \xi^{a'}_{\textrm S} &\equiv \Big[\Delta K^{a'}{}_b + L^{a'}{}_{bc} \xi^c + O(\boldsymbol{\xi}^2)\Big] \xi^b \nonumber \\
    &\hspace{2em}+ \left[\Upsilon^{a'}{}_b + \Psi^{a'}{}_{bc} \xi^c + O(\boldsymbol{\xi}^2)\right] s^b + O(\boldsymbol{s})^2, \displaybreak[0] \label{eqn:Delta_xi_decomp} \\
    \Delta p^{a'} &= m \frac{\uD}{\ud \tau_1} \Delta \xi^{a'}_{\textrm S} + O(\boldsymbol{s})^2 \displaybreak[0] \label{eqn:Delta_p_decomp} \\
    \Delta s^{a'} &\equiv \left[\Sigma^{a'}{}_{bc} \xi^c + O(\boldsymbol{\xi}^2, \boldsymbol{s})\right] s^b.
  \end{align}
\end{subequations}
The first of these expressions can be considered as definitions of $\Upsilon^{a'}{}_b$ and $\Psi^{a'}{}_{bc}$, and the third as a definition of $\Sigma^{a'}{}_{bc}$.
These are all bitensors determined by the curve $\gamma$ whose explicit forms will be calculated in Sec.~\ref{sec:spin_comp}.
The quantities $\Delta K^{a'}{}_b$ and $L^{a'}{}_{bc}$ are the same as those introduced in Sec.~\ref{sec:curve_deviation_intro}, and their values are given in Sec.~\ref{sec:curve_deviation}.
Equation~\eqref{eqn:Delta_p_decomp} is a result that will be proven in Sec.~\ref{sec:spin_comp}.
Our final expressions for these observables are given in Eq.~\eqref{eqn:spin_obs_final}; assuming weak curvature, we find that these expressions are given by

\begin{subequations} \label{eqn:spin_weak}
  \begin{align}
    \Sigma^\alpha{}_{\beta\gamma} &= -\int_{\tau_0}^{\tau_1} \ud \tau_2 R^\alpha{}_{\beta\gamma\delta} (\tau_2) \dot{\gamma}^\delta + O(\boldsymbol{R}^2), \\
    \Upsilon^\alpha{}_\beta &= -\int_{\tau_0}^{\tau_1} \ud \tau_2 \int_{\tau_0}^{\tau_2} \ud \tau_3 (R^*)^\alpha{}_{\gamma\beta\delta} (\tau_3) \dot{\gamma}^\gamma \dot{\gamma}^\delta + O(\boldsymbol{R}^2), \\
    \Psi^\alpha{}_{\beta\gamma} &= -\int_{\tau_0}^{\tau_1} \ud \tau_2 \int_{\tau_0}^{\tau_2} \ud \tau_3 [\nabla_\gamma (R^*)^\alpha{}_{\delta\beta\epsilon}] (\tau_3) \dot{\gamma}^\delta \dot{\gamma}^\epsilon \nonumber \\
    &\hspace{1em}+ O(\boldsymbol{R}^2).
  \end{align}
\end{subequations}
These expressions are much simpler than the corresponding expressions for the holonomy, or for curve deviation, primarily because the motion of the spinning body is already specified, which requires that the relative velocity and accelerations are given by very particular forms.

\subsection{Feasibility of measurement} \label{sec:feasibility}

All of the persistent observables in this paper are (in principle) measurable by some sort of detector, because these observables are defined in terms of a procedure that an observer can physically carry out.
For curve deviation and our observable involving a spinning test particle, these procedures are relatively straightforward to perform.
The former requires only a means of measuring separation and for the observers to keep track of their respective accelerations, while the latter only requires a method by which an observer can measure the momentum and spin of a particle in addition to the separation.
The holonomy observable is somewhat more complex, as it requires the two observers to measure the local curvature of spacetime (potentially by carrying around small gravitational wave detectors themselves).
The observers could then use the measured curvature to evolve the quantities $P^a$ and $J^{ab}$ according to Eq.~\eqref{eqn:kappa_transport}, and finally compare their results at the end.

A far simpler method that one could use to measure these persistent observables, without constructing new types of detectors, would be to take advantage of the fact that the values of these persistent observables can be written, in the weak curvature limit, in terms of integrals of the Riemann tensor (and its derivatives) along the worldline of one of the observers.
For observers far from an astrophysical source of gravitational waves, the weak-curvature limit is expected to be valid.
Moreover, when the observer is far enough from the source, the gravitational waves can be well approximated by plane waves, so the derivatives of the Riemann tensor can be expressed solely in terms of derivatives with respect to retarded time.
At fixed radius, retarded time is an affine parameter for the worldline of the observer, which allows terms involving integrals of the derivatives of the Riemann tensor to be written in terms of the Riemann tensor evaluated at the end points.
Gravitational wave detectors measure the components of the Riemann tensor, and these components can be integrated in time while the gravitational waves are passing by.
Having made these measurements of the integrated Riemann tensor, gravitational wave detectors could then use our weak curvature results to deduce what the value of any of the persistent observables in this paper would have been if the detector had, in fact, been carrying out the operations by which these observables are defined.

In this regime, and in the case where there is no acceleration, the weak curvature results in the preceding sections, that is, Eqs.~\eqref{eqn:curve_dev_weak},~\eqref{eqn:spin_weak}, and~\eqref{eqn:holonomy_lambda_weak} [when combined with Eq.~\eqref{eqn:holonomy_omega_weak}] involve only one, two, and three time integrals of the Riemann tensor along the worldline of the detector (allowing for acceleration terms, these results include more time integrals of the Riemann tensor).
Coincidentally, these numbers of time integrals of the Riemann tensor have appeared in previous discussions of persistent gravitational wave observables: one time integral for the relative proper time, velocity, and rotation observables [Eqs.~\eqref{eqn:time_mem},~\eqref{eqn:vel_mem}, and~\eqref{eqn:rot_mem}], two for the displacement memory [Eq.~\eqref{eqn:disp_mem}], and three for the subleading displacement memory [Eq.~\eqref{eqn:subdisp_mem}].
As such, in the limit discussed in this section, the only new information contained in these observables is the higher time integrals of the Riemann tensor that arise in the holonomy and curve deviation when there are acceleration terms.
In situations where this limit is not appropriate, such as general nonlinear gravitational wave spacetimes, the observables presented in this paper are not degenerate with those previously discussed in the literature.

\section{Review of techniques of covariant bitensors} \label{sec:memory_bitensor}

In this section, we provide a review of techniques that we will be using in this paper to compute the results provided in Sec.~\ref{sec:memory_comp}.
In the following three subsections, we follow the formalisms of~\cite{Flanagan2016} for introducing the idea of using a connection on a vector bundle to understand linear and angular momentum transport,~\cite{Poisson2004} for a brief review of bitensors, and~\cite{Vines2014b} for computations of holonomies.

\subsection{The linear and angular momentum bundle}

In this section, we use the notion of vector bundles, notably the idea of a \emph{direct-} (or \emph{Whitney-}) \emph{sum bundle} $E_1 \oplus E_2$ of two vector bundles $E_1$ and $E_2$.
This is the vector bundle obtained by taking, at each point in our manifold, the direct sum of the two vector spaces associated with $E_1$ and $E_2$ at that point.
As described in~\cite{Flanagan2016}, the ```linear and angular momentum bundle,'' which we call $\mathcal{A}$, is given by

\begin{equation}
  \mathcal{A} = T \mathcal{M} \oplus \Lambda^2 T \mathcal{M},
\end{equation}
where $T \mathcal{M}$ is the tangent bundle and  $\Lambda^2 T \mathcal{M}$ is the bundle of antisymmetric rank $(2, 0)$ tensors, or bivectors (dual two-forms).
We write a section of this bundle as

\begin{equation} \label{eqn:bundle_def}
  X^A = \begin{pmatrix}
    P^a \\
    J^{ab}
  \end{pmatrix},
\end{equation}
for some tensor fields $P^a$ and $J^{ab}$.
For any quantity that is part of a matrix on this vector bundle, we denote the various components as follows:

\begin{equation} \label{eqn:bundle_mat}
  A^A{}_C \equiv \begin{pmatrix}
    \smallunderset{PP}{A}{}^a{}_c & \smallunderset{PJ}{A}{}^a{}_{cd} \vspace{0.5em} \\
    \smallunderset{JP}{A}{}^{ab}{}_c & \smallunderset{JJ}{A}{}^{ab}{}_{cd}
  \end{pmatrix}.
\end{equation}
This is exactly how the various components of the holonomy were denoted in Eq.~\eqref{eqn:holonomy_vars}.

There is furthermore the notion of a connection on this bundle.
The connection that we are typically concerned with is given by rewriting Eq.~\eqref{eqn:kappa_transport} as

\begin{equation} \label{eqn:kappa_transport_bundle}
  0 = k^a \vkappa{\nabla}_a X^B \equiv k^a \nabla_a X^B + k^a \vkappa{C}{}^B{}_{Ca} X^C,
\end{equation}
where $\nabla_a$ denotes the natural extension of the Levi-Civita connection to the angular momentum bundle, and

\begin{equation} \label{eqn:kappa_connection_coefficients}
  \vkappa{C}{}^A{}_{Ce} = \begin{pmatrix}
    0 & \vkappa{K}{}^a{}_{ecd} \\
    2 \delta^{[a}{}_e \delta^{b]}{}_c & 0
  \end{pmatrix}.
\end{equation}

Most of what follows in this section can be applied to any connection, so we proceed in full generality, using $\hat{\nabla}_a$ and $\check{\nabla}_a$ as arbitrary connections.
We will also use capital Latin indices for indices on a generic bundle.
Throughout this paper we add a diacritical mark above the core symbol of any tensor that depends on a given connection with the same diacritic above the connection; for example, the parallel propagator $\hat{g}^{A'}{}_A$ [Eq.~\eqref{eqn:prop_def}] is defined with respect to the connection $\hat{\nabla}_a$.
Furthermore, for quantities that depend on two connections, we add both diacritics above the core symbol for the diacritical marks associated with the two connections; for example, the connection coefficient $\hat{\check C}^A{}_{Bc}$ [Eq.~\eqref{eqn:C_def}] is defined with respect to $\hat{\nabla}_a$ and $\check{\nabla}_a$.

\subsection{Bitensors on vector bundles} \label{sec:bitensors}

In this section we define the \emph{parallel propagators} $\pb{\gamma} \hat{g}^{A'}{}_A$ and $\pb{\gamma} \hat{g}^A{}_{A'}$, which are bitensors at $x \equiv \gamma(\tau)$ and $x' \equiv \gamma(\tau')$, and are defined with respect to a connection $\hat{\nabla}_a$ on some arbitrary vector bundle on our manifold.
We then find expressions to relate parallel propagators that are defined with respect to different connections.

To construct the parallel propagator, consider a basis of vectors at $x = \gamma(\tau)$ denoted by $\{(\hat{e}_\Gamma)^A \mid \Gamma = 1, \ldots, d\}$ and the basis of one-forms dual to this basis denoted by $\{(\hat{\omega}^\Gamma)_A \mid \Gamma = 1, \ldots, d\}$.
Now, parallel transport both $(\hat{e}_\Gamma)^A$ and $(\hat{\omega}^\Gamma)_A$ along $\gamma$ from $\gamma(\tau)$ to $\gamma(\tau')$, with respect to the connection $\hat{\nabla}_a$, to yield $(\hat{e}_\Gamma)^{A'}$ and $(\hat{\omega}^\Gamma)_{A'}$.
From these tetrads, we can define the parallel propagators by

\begin{equation} \label{eqn:prop_def}
  \pb{\gamma} \hat{g}^{A'}{}_A \equiv \sum_{\Gamma = 1}^d (\hat{e}_\Gamma)^{A'} (\hat{\omega}^\Gamma)_A,
\end{equation}
with the parallel propagator $\pb{\gamma} \hat{g}^A{}_{A'}$ defined by switching $A$ and $A'$.
Note that this is not the usual parallel propagator defined in, say,~\cite{Poisson2004}, as the bases are parallel transported along a specific curve.
This is the significance of the subscripted $\gamma$ that is added to the left of the $g$.
Moreover, this definition allows for connections that are not metric compatible, and does not require these bases to be either orthogonal or normalized with respect to any metric.

As the bases were parallel transported with respect to $\hat{\nabla}_a$, the parallel propagators satisfy

\begin{equation}
  \dot{\gamma}^b \hat{\nabla}_b \pb{\gamma} \hat{g}^{A'}{}_A = 0, \qquad \dot{\gamma}^{b'} \hat{\nabla}_{b'} \pb{\gamma} \hat{g}^{A'}{}_A = 0.
\end{equation}
A similar result holds for $\pb{\gamma} \hat{g}^A{}_{A'}$.
Note that this means that $Y^{A'} (\tau) \equiv X^A \pb{\gamma} \hat{g}{}^{A'}{}_A$ is the unique solution to the differential equation

\begin{equation}
  \dot{\gamma}^{b'} \hat{\nabla}_{b'} Y^{A'} (\tau) = 0,
\end{equation}
with boundary condition $Y^A (\tau) = X^A$.
A similar result holds for $Y_{A'} (\tau) \equiv X_A \pb{\gamma} \hat{g}{}^A{}_{A'}$.
Moreover, one can show that

\begin{equation} \label{eqn:prop_inverse}
  \pb{\gamma} \hat{g}^A{}_{A'} \pb{\gamma} \hat{g}^{A'}{}_B = \delta^A{}_B,
\end{equation}
as well as the same result with primed and unprimed indices switched.

Now, consider two points in a \emph{convex normal neighborhood}, that is, in a small enough region such that there is a unique geodesic $\Gamma_{(x, x')}$ satisfying

\begin{equation} \label{eqn:Gamma_def}
  \Gamma_{(x, x')} (0) = x, \qquad \Gamma_{(x, x')} (1) = x'.
\end{equation}
In this neighborhood, we can define the other bitensor which we will be using, Synge's world function $\sigma(x, x')$, as half of the squared distance along $\Gamma_{(x, x')}$:

\begin{equation}
  \sigma(x, x') = \frac{1}{2} \int_0^1 \ud \lambda\; g_{a''b''} \dot{\Gamma}_{(x, x')}^{a''} \dot{\Gamma}_{(x, x')}^{b''},
\end{equation}
where $x'' \equiv \Gamma_{(x, x')} (\lambda)$.
The derivatives of Synge's world function are denoted by appending indices onto $\sigma(x, x')$:

\begin{equation} \label{eqn:synge_diff_def}
  \nabla_{a_1} \cdots \nabla_{a_n} \sigma (x, x') \equiv \sigma_{a_n \cdots a_1} (x, x').
\end{equation}
These indices can be indices at either $x$ or $x'$.
Note that (as shown, for example, in~\cite{Poisson2004})

\begin{equation} \label{eqn:sep_vec}
  \sigma^a (x') = -\dot{\Gamma}_{(x, x')}^a, \qquad \sigma^{a'} (x) = \dot{\Gamma}_{(x, x')}^{a'}.
\end{equation}
This shall be our primary use for Synge's world function, since it provides a notion of separation vector between two nearby points.

In a convex normal neighborhood, we can also define the usual parallel propagator $\hat{g}^{A'}{}_A$ by

\begin{equation} \label{eqn:usual_prop}
  \hat{g}^{A'}{}_A \equiv \pb{\Gamma_{(x, x')}} \hat{g}^{A'}{}_A.
\end{equation}
This bitensor satisfies

\begin{equation} \label{eqn:usual_prop_prop}
  \sigma^b (x') \hat{\nabla}_b \hat{g}^{A'}{}_A = \sigma^{b'} (x) \hat{\nabla}_{b'} \hat{g}^{A'}{}_A = 0,
\end{equation}
with again a similar result holding for $\hat{g}^A{}_{A'}$.

Now, consider the case where we have two connections, $\hat{\nabla}_a$ and $\check{\nabla}_a$, defined on this vector bundle.
Define

\begin{equation} \label{eqn:C_def}
  (\hat{\nabla}_b - \check{\nabla}_b) X^A \equiv \hat{\check C}^A{}_{Cb} X^C,
\end{equation}
where clearly $\hat{\check C}^A{}_{Bc} = -\check{\hat C}^A{}_{Bc}$.
Note that $\pb{\gamma} \hat{g}{}^{A'}{}_A$ satisfies

\begin{equation} \label{eqn:U_diff}
  \frac{\ud}{\ud \tau'} \left(\pb{\gamma} \check{g}^A{}_{A'} \pb{\gamma} \hat{g}^{A'}{}_B\right) = -\pb{\gamma} \check{g}^A{}_{A'} \hat{\check C}^{A'}{}_{C'd'} \dot{\gamma}^{d'} \pb{\gamma} \hat{g}^{C'}{}_B.
\end{equation}
Thus, $\pb{\gamma} \hat{g}^{A'}{}_A$ is a solution to the following integral equation:

\begin{equation} \label{eqn:U_int}
  \pb{\gamma} \hat{g}^{A'}{}_B = \pb{\gamma} \check{g}^{A'}{}_A \left(\delta^A{}_B - \int_\tau^{\tau'} \ud \tau'' \pb{\gamma} \hat{\check A}^A{}_{B''} \pb{\gamma} \hat{g}^{B''}{}_B\right),
\end{equation}
where $x'' \equiv \gamma(\tau'')$ and

\begin{equation}
  \pb{\gamma} \hat{\check A}^A{}_{B'} \equiv \pb{\gamma} \check{g}^A{}_{A'} \hat{\check C}^{A'}{}_{B'c'} \dot{\gamma}^{c'}.
\end{equation}
By the same logic, we can show that

\begin{equation} \label{eqn:U_inv_int}
  \pb{\gamma} \hat{g}^A{}_{B'} = \pb{\gamma} \check{g}^A{}_{A'} \left(\delta^{A'}{}_{B'} + \int_\tau^{\tau'} \ud \tau'' \pb{\gamma} \hat{\check A}^{A'}{}_{B''} \pb{\gamma} \hat{g}^{B''}{}_{B'}\right),
\end{equation}
We typically solve Eqs.~\eqref{eqn:U_int} and~\eqref{eqn:U_inv_int} iteratively, either by truncating the expansion based on a particular approximation scheme, or by exploiting the fact that $\pb{\gamma} \hat{\check A}^A{}_{B'}$ is nilpotent in some circumstances.

With these basic bitensors defined, we now consider holonomies.

\subsection{Holonomies of transport laws}

In terms of bitensors, the holonomy of a connection $\hat{\nabla}_a$ around some closed curve $\mathcal{C}$ is given by

\begin{equation} \label{eqn:holo_def}
  \pb{\mathcal{C}} \hat{\Lambda}^A{}_B = \pb{\mathcal{C}} \hat{g}^A{}_B.
\end{equation}
If the closed curve is only piecewise smooth, composed of smooth paths $\mathcal{P}_1, \ldots, \mathcal{P}_n$ with end points $x', \ldots, x^{(n)}$, then we write

\begin{equation} \label{eqn:holo_decomp}
  \pb{\mathcal{C}} \hat{\Lambda}^A{}_B = \pb{\mathcal{P}_n} \hat{g}^A{}_{B^{(n)}} \cdots \pb{\mathcal{P}_1} \hat{g}^{B'}{}_B
\end{equation}
In the next few sections, we will find expressions for the holonomies for various shapes, for an arbitrary connection $\hat{\nabla}_a$.

\subsubsection{Nongeodesic polygons}

\begin{figure}[t!]
  \centering
  \includegraphics{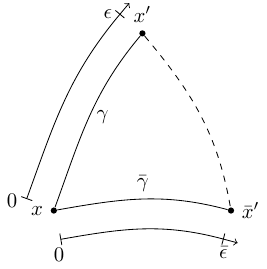}
  \caption{\label{fig:triangle} A nongeodesic triangle (generalizing Fig.~3 of~\cite{Vines2014b}), where the two sides $\gamma$ and $\bar{\gamma}$ are arbitrary curves (with affine parameter lengths $\epsilon$ and $\bar{\epsilon}$), and where the third side is formed by joining the two end points by the unique geodesic extending between them.}
\end{figure}

First, following~\cite{Vines2014b}, we show that the holonomy around a (nongeodesic) triangle is given by expressions involving the Riemann tensor associated with the connection on the vector bundle.
Explicitly, consider the triangle depicted in Fig.~\ref{fig:triangle}, where two edges are segments of arbitrary curves $\gamma$ and $\bar{\gamma}$ that meet at a point $x \equiv \gamma(0) \equiv \bar{\gamma} (0)$.
Join $x' \equiv \gamma(\epsilon)$ and $\bar{x}' \equiv \bar{\gamma} (\bar{\epsilon})$ by the unique geodesic between them.

Now, the holonomy around this triangle is given by

\begin{equation} \label{eqn:holo_triangle_def}
  \pb{\triangle} \hat{\Lambda}^A{}_B (\gamma, \bar{\gamma}; \epsilon, \bar{\epsilon}) \equiv \pb{\gamma} \hat{g}^A{}_{\bar A'} \hat{g}^{\bar A'}{}_{B'} \pb{\gamma} \hat{g}^{B'}{}_B.
\end{equation}
Expanding the holonomy in a Taylor series in $\epsilon$ and $\bar{\epsilon}$ as

\begin{equation} \label{eqn:holo_triangle_exp}
  \pb{\triangle} \hat{\Lambda}^A{}_B (\gamma, \bar{\gamma}; \epsilon, \bar{\epsilon}) \equiv \sum_{m, n = 0}^\infty \frac{\epsilon^m \bar{\epsilon}^n}{m! n!} \pb{\triangle} \underset{m, n}{\hat{\Lambda}}{}^A{}_B (\gamma, \bar{\gamma}),
\end{equation}
we find that

\begin{equation} \label{eqn:holo_triangle_series}
  \begin{split}
    \pb{\triangle} \underset{m, n}{\hat{\Lambda}}{}^A{}_B (\gamma, \bar{\gamma}) &= \left.\frac{\partial^{m + n}}{\partial \epsilon^m \partial \bar{\epsilon}^n} \Lambda^A{}_B (\gamma, \bar{\gamma}; \epsilon, \bar{\epsilon}) \right|_{\epsilon = 0, \bar{\epsilon} = 0} \\
    &= \left[\left(\dot{\gamma}^{c'} \hat{\nabla}_{c'}\right)^m \left(\dot{\bar \gamma}^d \hat{\nabla}_d\right)^n \hat{g}^A{}_{B'}\right]_{x' \to x}. \\
  \end{split}
\end{equation}
The brackets denote \emph{coincidence limits}, which are limits of a bitensorial expression as one of the points approaches the other.
A brief review of coincidence limits is given in Appendix~\ref{sec:coincidence}, which is based on parts of Poisson's review article~\cite{Poisson2004}.
Keeping terms to only quadratic order yields [using the expressions for the coincidence limits of parallel propagators from Eq.~\eqref{eqn:prop_coincidence}]

\begin{equation} \label{eqn:holo_triangle_quadratic}
  \pb{\triangle} \hat{\Lambda}^A{}_B (\gamma, \bar{\gamma}; \epsilon, \bar{\epsilon}) = \delta^A{}_B - \frac{1}{2} \epsilon \bar{\epsilon} \dot{\gamma}^c \dot{\bar \gamma}^d \hat{R}^A{}_{Bcd} + O(\epsilon, \bar{\epsilon})^3,
\end{equation}
where $\hat{R}^A{}_{Bcd}$ is the curvature tensor defined with respect to the connection $\hat{\nabla}_a$ and is defined by

\begin{equation} \label{eqn:R_def}
  2 \hat{\nabla}_{[c} \hat{\nabla}_{d]} X^A \equiv \hat{R}^A{}_{Bcd} X^B.
\end{equation}
For two connections $\hat{\nabla}_a$ and $\check{\nabla}_a$, their curvature tensors are related by

\begin{equation} \label{eqn:R_diff}
  \hat{R}^A{}_{Bcd} = \check{R}^A{}_{Bcd} + 2 \check{\nabla}_{[c} \hat{\check C}^A{}_{|B|d]} + 2 \hat{\check C}^A{}_{E[c} \hat{\check C}^E{}_{|B|d]}.
\end{equation}
Note that Eq.~\eqref{eqn:holo_triangle_quadratic} does not contain any acceleration terms at this order; moreover, it reduces to the results of~\cite{Vines2014b} for the metric-compatible connection on the tangent bundle.

\begin{figure}[b!]
  \centering
  \includegraphics{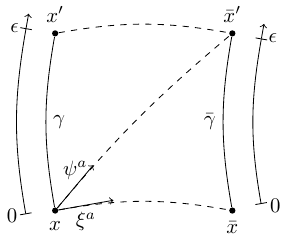}
  \caption{\label{fig:square} A nongeodesic square, with two sides $\gamma$ and $\bar{\gamma}$ that are arbitrary curves (with equal affine parameter length $\epsilon$), and where the other two sides are the unique geodesics between the two initial and final points, respectively.
    A third unique geodesic forms the diagonal.
    We denote the tangents (normalized such that the total affine parameter lengths are 1) to these unique geodesics at $x$ by $\xi^a$ and $\psi^a$.}
\end{figure}

Now, we consider the holonomy around a square, such as that given in Fig.~\ref{fig:square}: this square is determined by two arbitrary curves $\gamma$ and $\bar{\gamma}$, with the pairs of initial and final points, respectively, connected by the unique geodesics between them.
The initial points are labeled $x = \gamma(0)$ and $\bar{x} = \bar{\gamma} (0)$, and the final points are labeled $x' = \gamma(\epsilon)$ and $\bar{x}' = \bar{\gamma} (\epsilon)$, and we assume that $\epsilon$ is small.
We define two ``separation vectors''

\begin{equation} \label{eqn:square_separation_def}
  \xi^a (\bar{x}) \equiv -\sigma^a (\bar{x}) = g^a{}_{\bar a} \sigma^{\bar a} (x), \qquad \psi^a (\bar{x}, \epsilon) \equiv -\sigma^a (\bar{x}').
\end{equation}
In terms of these quantities, the holonomy around this square is given by

\begin{widetext}
\begin{equation} \label{eqn:holo_square_def}
  \begin{split}
    \pb{\square} \hat{\Lambda}^A{}_B (\gamma, \bar{\gamma}; \epsilon) &\equiv \hat{g}^A{}_{\bar A} \pb{\triangle} \left(\hat{\Lambda}^{-1}\right){}^{\bar A}{}_{\bar C} (\bar{\gamma}, \Gamma_{(\bar{x}, x)}; \epsilon, 1) \hat{g}^{\bar C}{}_C \pb{\triangle} \hat{\Lambda}^C{}_B (\gamma, \Gamma_{(x, \bar{x}')}; \epsilon, 1) \\
    &= \delta^A{}_B - \frac{\epsilon}{2} \bigg[\dot{\gamma}^c \psi^d (\bar{x}, \epsilon) \hat{R}^A{}_{Bcd} + \hat{g}^A{}_{\bar A} \hat{g}^{\bar B}{}_B \dot{\bar \gamma}^{\bar c} g^{\bar d}{}_d \xi^d (\bar x) \hat{R}^{\bar A}{}_{\bar B \bar c \bar d} + O(\boldsymbol{\xi}, \boldsymbol{\psi})^2\bigg] + O(\epsilon^2).
  \end{split}
\end{equation}
\end{widetext}
Note that we have traversed this loop in a way such that we can use Eq.~\eqref{eqn:holo_triangle_quadratic}, which was only established with two of the sides of the triangle being nongeodesic.
First, note that the second term within square brackets in Eq.~\eqref{eqn:holo_square_def} has the following coincidence limit:

\begin{equation}
  \hat{g}^A{}_{\bar A} \hat{g}^{\bar B}{}_B \hat{R}^{\bar A}{}_{\bar B \bar c \bar d} = g^c{}_{\bar c} g^d{}_{\bar d} \hat{R}^A{}_{Bcd} + O(\boldsymbol{\xi}). \label{eqn:R_exp} \\
\end{equation}
Next, the expansion for $\psi^a (\bar{x}, \epsilon)$ to lowest order in $\epsilon$ can be derived by noting that $\psi^a (\bar{x}, 0) = \xi^a (\bar{x})$:

\begin{equation} \label{eqn:psi_exp}
  \psi^a (\bar{x}, \epsilon)= \xi^a (\bar{x}) + O(\epsilon).
\end{equation}
Plugging these expressions into Eq.~\eqref{eqn:holo_square_def} gives

\begin{equation} \label{eqn:square_holonomy_quadratic}
  \begin{split}
    \pb{\square} \hat{\Lambda}^A{}_B (\gamma, \bar{\gamma}; \epsilon) = \delta^A{}_B &- \frac{\epsilon}{2} \left(\dot{\gamma}^c + g^c{}_{\bar c} \dot{\bar \gamma}^{\bar c}\right) \xi^d (\bar{x}) \hat{R}^A{}_{Bcd} \\
    &+ O(\epsilon^2, \boldsymbol{\xi}^2).
  \end{split}
\end{equation}

\subsubsection{Narrow loops} \label{sec:narrow}

\begin{figure}[t!]
  \centering
  \includegraphics{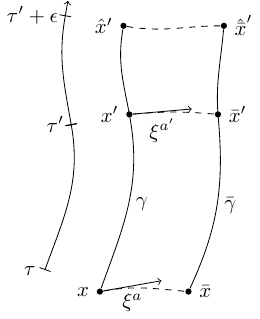}
  \caption{\label{fig:loop} Two nearby worldlines $\gamma$ and $\bar{\gamma}$, joined by unique geodesics between the start and end points of $\gamma$ and $\bar\gamma$, respectively.
    The separation vector between the two worldlines is denoted by $\xi^a$.}
\end{figure}

Finally, consider the holonomy about the loop in Fig.~\ref{fig:loop}, which we will denote by $\hat{\Lambda}^A{}_B (\gamma, \bar{\gamma}; \tau')$.
This is a curve defined by two nearby, timelike, and affine-parametrized curves $\gamma$ and $\bar{\gamma}$, which are connected at the points $\gamma(\tau)$ and $\bar{\gamma} (\tau)$, as well as the points $\gamma(\tau')$ and $\bar{\gamma} (\tau')$.
Here, we are explicitly using the isochronous correspondence mentioned in Sec.~\ref{sec:curve_deviation_intro} above (see~\cite{Vines2014a} for a review; we follow their conventions and general argument), where the separation vector connects points with equal values of affine parameter.
For convenience, we assume that this shared affine parameter is the proper time of both worldlines, and thus is fixed up to an additive constant (which can be set initially by requiring $\xi^a \dot{\gamma}_a = 0$, for example).

The separation vector $\xi^a (\bar{x})$ in Eq.~\eqref{eqn:square_separation_def} is now a function of proper time along one of the worldlines, which we will simply denote by

\begin{equation} \label{eqn:separation_def}
  \xi^a \equiv -\sigma^a [\bar{\gamma} (\tau)],
\end{equation}
for any $\tau$ along the worldline (note that, as usual, we suppress the dependence of the derivative of Synge's world function on $\gamma(\tau)$, since that is apparent from the index).
We further define

\begin{equation} \label{eqn:dotxi_def}
  \dot{\xi}^a \equiv \left(\dot{\gamma}^b \nabla_b + \dot{\bar{\gamma}}^{\bar b} \nabla_{\bar b}\right ) \xi^a,
\end{equation}
which yields, from Eq.~\eqref{eqn:synge_coincidence},

\begin{equation}
  \dot{\xi}^a = g^a{}_{\bar a} \dot{\bar \gamma}^{\bar a} - \dot{\gamma}^a + \frac{1}{6} R^a{}_{cbd} \xi^c \xi^d (g^b{}_{\bar b} \dot{\bar \gamma}^{\bar b} + 2 \dot{\gamma}^b) + O(\boldsymbol{\xi}^3).
\end{equation}
Inverting this equation to first order, and then plugging in that solution to invert it to second order, we find that

\begin{equation} \label{eqn:utilde}
  \begin{split}
    \dot{\bar{\gamma}}^{\bar a} = &g^{\bar a}{}_a \left[\dot{\gamma}^a + \dot{\xi}^a - \frac{1}{6} R^a{}_{bcd} \xi^b \left(3 \dot{\gamma}^c + \dot{\xi}^c\right) \xi^d\right] \\
    &+ O(\boldsymbol{\xi}^3).
  \end{split}
\end{equation}

We now continue with the calculation of the holonomy.
For any $\tau'$ along the worldline, and a given $\epsilon$, we have

\begin{equation} \label{eqn:holo_narrow_def}
  \begin{split}
    \hat{\Lambda}^A{}_B (\gamma, \bar{\gamma}; \tau' + \epsilon) &= \hat{\Lambda}^A{}_C (\gamma, \bar{\gamma}; \tau') \\
    &\hspace{1em}\times \pb{\gamma} \hat{g}^C{}_{C'} \pb{\Box} \hat{\Lambda}^{C'}{}_{D'} (\gamma, \bar{\gamma}; \epsilon) \pb{\gamma} \hat{g}^{D'}{}_B.
  \end{split}
\end{equation}
Taking the limit $\epsilon \to 0$, we find the following differential equation for the holonomy of a narrow loop:

\begin{equation} \label{eqn:holo_narrow_diff}
  \begin{split}
    \frac{\ud}{\ud \tau'} &\hat{\Lambda}^A{}_B (\gamma, \bar{\gamma}; \tau') \\
    &= \hat{\Lambda}^A{}_E (\gamma, \bar{\gamma}; \tau') \pb{\gamma} \hat{g}^E{}_{E'} \pb{\gamma} \hat{g}^{B'}{}_B \hat{R}^{E'}{}_{B'c'd'} \xi^{c'} \dot{\gamma}^{d'} \\
    &\hspace{1em}+ O(\boldsymbol{\xi}, \dot{\boldsymbol \xi})^2.
  \end{split}
\end{equation}
This differential equation can be solved iteratively.
Keeping terms at first order in $\xi^a$ and $\dot{\xi}^a$, we find

\begin{equation} \label{eqn:holo_narrow_soln}
  \begin{split}
    \hat{\Lambda}^A{}_B (\gamma, \bar{\gamma}; \tau') &- \delta^A{}_B \\
    &= \int_\tau^{\tau'} \ud \tau'' \pb{\gamma} \hat{g}^A{}_{A''} \hat{R}^{A''}{}_{B''c''d''} \xi^{c''} \dot{\gamma}^{d''} \pb{\gamma} \hat{g}^{B''}{}_B \\
    &\hspace{1em}+ O(\boldsymbol{\xi}, \dot{\boldsymbol \xi})^2.
  \end{split}
\end{equation}
This concludes our discussion of the holonomies of arbitrary connections, and we turn to applications of these results for calculations of persistent observables.

\section{Computations of persistent observables} \label{sec:memory_comp}

In this section, we provide explicit expressions for all of the persistent observables in Sec.~\ref{sec:memory_overview}, which we give using the formalism of covariant bitensors reviewed in the previous section.
In a subsequent paper in this series, we will give explicit expressions that are valid in particular spacetimes, essentially by determining the values of the so-called \emph{fundamental bitensors} in terms of which the final results of this section are given.
Results that are valid, assuming weak curvature, were given in Sec.~\ref{sec:memory_overview}.

Throughout this section, we use the convention that $x \equiv \gamma(\tau_0)$, $\bar{x} \equiv \bar{\gamma} (\tau_0)$, and in general,

\begin{equation}
  x^{(n)} \equiv \gamma(\tau_n), \qquad \bar{x}^{(n)} \equiv \bar{\gamma} (\tau_n),
\end{equation}
where $x^{(n)}$ is $x$ with $n$ primes.
When considering some arbitrary $\tau$, we also use $x$ and $\bar{x}$ for convenience.

\subsection{Curve deviation} \label{sec:curve_deviation}

The original memory observable considered in~\cite{Zeldovich1974} was based on the evolution of the separation vector between two nearby geodesics.
In this section, we review the computation of this separation vector in terms of the initial separation and its derivative, as well as the accelerations of the worldlines.
This forms the basis of the curve deviation observable introduced in Sec.~\ref{sec:curve_deviation_intro}.
By Eq.~\eqref{eqn:holo_narrow_soln}, this is also necessary to calculate the holonomy, as well as the persistent observable involving a spinning particle (as we will discuss in Sec.~\ref{sec:spin_comp}).
We carry out this calculation to second order in $\xi^a$ and $\dot{\xi}^a$.

To begin, we take another derivative of Eq.~\eqref{eqn:utilde}:

\begin{widetext}
\begin{equation} \label{eqn:atilde}
  \begin{split}
    \ddot{\bar{\gamma}}^{\bar a} = &\left(\dot{\gamma}^a + \dot{\xi}^a - \frac{1}{2} R^a{}_{bcd} \xi^b \dot{\gamma}^c \xi^d\right) \left(\dot{\gamma}^e \nabla_e + \dot{\bar \gamma}^{\bar e} \nabla_{\bar e}\right) g^{\bar a}{}_a \\
    &+ g^{\bar a}{}_a \left[\left(\delta^a{}_b - \frac{1}{2} R^a{}_{cbd} \xi^c \xi^d\right) \ddot{\gamma}^b + \left(\delta^a{}_b - \frac{1}{6} R^a{}_{cbd} \xi^c \xi^d\right) \ddot{\xi}^b - \frac{1}{2} \left(\xi^d \dot{\gamma}^e \nabla_e R^a{}_{bcd} + 2 \dot{\xi}^d R^a{}_{(b|c|d)}\right) \xi^b \dot{\gamma}^c\right] \\
    &+ O(\boldsymbol{\xi}, \dot{\boldsymbol \xi})^3. \\
  \end{split}
\end{equation}
Using the coincidence limits of derivatives of the parallel propagators in Eq.~\eqref{eqn:prop_coincidence}, we have that

\begin{equation}
  \left(\dot{\gamma}^b \nabla_b + \dot{\bar \gamma}^{\bar b} \nabla_{\bar b}\right) g^{\bar a}{}_a = g^{\bar a}{}_c \left[R^c{}_{abd} \left(\dot{\gamma}^d + \frac{1}{2} \dot{\xi}^d\right) + \frac{1}{2} \dot{\gamma}^d \xi^e \nabla_e R^c{}_{abd}\right] \xi^b + O(\boldsymbol{\xi}, \dot{\boldsymbol \xi})^3 .
\end{equation}
Thus, Eq.~\eqref{eqn:atilde} can be written as

\begin{equation}
  \begin{split}
    \ddot{\bar{\gamma}}^{\bar a} = g^{\bar a}{}_a \bigg\{&\left(\delta^a{}_b - \frac{1}{2} R^a{}_{cbd} \xi^c \xi^d\right) \ddot{\gamma}^b + \left(\delta^a{}_b - \frac{1}{6} R^a{}_{cbd} \xi^c \xi^d\right) \ddot{\xi}^b \\
    &+ \left[R^a{}_{cbd} \left(\dot{\gamma}^c \dot{\gamma}^d + \frac{1}{2} \dot{\gamma}^c \dot{\xi}^d + \dot{\xi}^c \dot{\gamma}^d\right) + \frac{1}{2} \dot{\gamma}^c \dot{\gamma}^d \xi^e \nabla_e R^a{}_{cbd} - \frac{1}{2} \dot{\gamma}^c \xi^d \dot{\gamma}^e \nabla_e R^a{}_{bcd} - \dot{\gamma}^c \dot{\xi}^d R^a{}_{(b|c|d)}\right] \xi^b\bigg\} \\
    &+ O(\boldsymbol{\xi}, \dot{\boldsymbol \xi})^3, \\
  \end{split}
\end{equation}
which can be solved for $\ddot\xi^a$:

\begin{equation} \label{eqn:ddot_xi}
  \begin{split}
    \ddot{\xi}^a = &- R^a{}_{cbd} \dot{\gamma}^c \dot{\gamma}^d \xi^b - 2 R^a{}_{cbd} \xi^b \dot{\xi}^c \dot{\gamma}^d - \nabla_{(e} R^a{}_{c)bd} \xi^b \xi^c \dot{\gamma}^d \dot{\gamma}^e \\
    &+ \left(\delta^a{}_b + \frac{1}{6} R^a{}_{cbd} \xi^c \xi^d\right) g^b{}_{\bar b} \ddot{\bar \gamma}^{\bar b} - \left(\delta^a{}_b - \frac{1}{3} R^a{}_{cbd} \xi^c \xi^d\right) \ddot{\gamma}^b + O(\boldsymbol{\xi}, \dot{\boldsymbol \xi})^3.
  \end{split}
\end{equation}
\end{widetext}

To solve this differential equation, we first define the solutions to its homogeneous, linearized version, which are given by

\begin{equation}
  \xi^{a'} = \pb{\gamma} K^{a'}{}_a \xi^a + (\tau_1 - \tau_0) \pb{\gamma} H^{a'}{}_a \dot{\xi}^a,
\end{equation}
where $\pb{\gamma} K^{a'}{}_a$ and $\pb{\gamma} H^{a'}{}_a$ are known as \emph{Jacobi propagators} (as defined in, for example,~\cite{Vines2014a}).
It is conventional not to absorb the factor of $\tau_1 - \tau_0$ into the definition of $\pb{\gamma} H^{a'}{}_a$, as it is convenient for defining the Jacobi propagators in terms of Synge's world function.
Here, note that we are instead defining $\pb{\gamma} K^{a'}{}_a$ and $(\tau_1 - \tau_0) \pb{\gamma} H^{a'}{}_a$ to be the solutions to the equation

\begin{equation} \label{eqn:jacobi}
  \frac{\uD^2}{\ud \tau_1^2} A^{a'}{}_a = -R^{a'}{}_{c'b'd'} \dot{\gamma}^{c'} \dot{\gamma}^{d'} A^{b'}{}_a,
\end{equation}
with boundary conditions

\begin{subequations} \label{eqn:K_H_bc}
  \begin{align}
    \left.\pb{\gamma} K^{a'}{}_b\right|_{\tau_1 = \tau_0} &= \left.\frac{\uD}{\ud \tau_1} \left[(\tau_1 - \tau_0) \pb{\gamma} H^{a'}{}_b\right]\right|_{\tau_1 = \tau_0} = \delta^a{}_b, \displaybreak[0] \\
    \left.\frac{\uD}{\ud \tau_1} \pb{\gamma} K^{a'}{}_b\right|_{\tau_1 = \tau_0} &= \left.(\tau_1 - \tau_0) \pb{\gamma} H^{a'}{}_b\right|_{\tau_1 = \tau_0} = 0.
  \end{align}
\end{subequations}

To solve Eq.~\eqref{eqn:ddot_xi} to second order, we note that we can insert the linear solution into this nonlinear equation, which now becomes an inhomogeneous, linear differential equation with a source term:

\begin{equation} \label{eqn:deviation_source}
  \begin{split}
    \ddot{\xi}^{a'} = &-R^{a'}{}_{c'b'd'} \dot{\gamma}^{c'} \dot{\gamma}^{d'} \xi^{b'} + S^{a'} [\boldsymbol{\xi}, \dot{\boldsymbol{\xi}}, \ddot{\boldsymbol{\gamma}}, \ddot{\boldsymbol{\bar \gamma}}] \\
    &+ O(\boldsymbol{\xi}, \dot{\boldsymbol \xi})^3,
  \end{split}
\end{equation}
where $S^{a'} [\boldsymbol{\xi}, \dot{\boldsymbol{\xi}}, \ddot{\boldsymbol{\gamma}}, \ddot{\boldsymbol{\bar \gamma}}]$ is a function of the initial $\xi^a$ and $\dot{\xi}^a$.
The solution to this equation, valid to second order, is given by

\begin{equation} \label{eqn:x_soln}
  \begin{split}
    \xi^{a'} = &\pb{\gamma} K^{a'}{}_a \xi^a + (\tau_1 - \tau_0) \pb{\gamma} H^{a'}{}_a \dot{\xi}^a \\
    &+ \int_{\tau_0}^{\tau_1} \ud \tau_2 (\tau_1 - \tau_2) \pb{\gamma} H^{a'}{}_{a''} S^{a''} [\boldsymbol{\xi}, \dot{\boldsymbol{\xi}}, \ddot{\boldsymbol{\gamma}}, \ddot{\boldsymbol{\bar \gamma}}] \\
    &+ O(\boldsymbol{\xi}, \dot{\boldsymbol \xi})^3.
  \end{split}
\end{equation}
For brevity, we merely check whether the solution in Eq.~\eqref{eqn:x_soln}, which can be derived using techniques similar to those in Sec.~\ref{sec:bitensors}, satisfies Eq.~\eqref{eqn:deviation_source}.
To verify this solution, note that, applying the Leibniz integral rule twice, we see

\begin{widetext}
\begin{equation}
  \begin{split}
    \frac{\uD^2}{\ud \tau_1^2} \int_{\tau_0}^{\tau_1} \ud \tau_2 (\tau_1 - \tau_2) \pb{\gamma} H^{a'}{}_{a''} S^{a''} [\boldsymbol{\xi}, \dot{\boldsymbol{\xi}}, \ddot{\boldsymbol{\gamma}}, \ddot{\boldsymbol{\bar \gamma}}] &= \left.\frac{\uD}{\ud \tau_1} \Big[(\tau_1 - \tau_2) \pb{\gamma} H^{a'}{}_{a''}\Big] S^{a''} [\boldsymbol{\xi}, \dot{\boldsymbol{\xi}}, \ddot{\boldsymbol{\gamma}}, \ddot{\boldsymbol{\bar \gamma}}]\right|_{\tau_2 = \tau_1} \\
    &\hspace{1em}+ \int_{\tau_0}^{\tau_1} \ud \tau_2 \frac{\uD^2}{\ud \tau_1^2} \Big[(\tau_1 - \tau_2) \pb{\gamma} H^{a'}{}_{a''}\Big] S^{a''} [\boldsymbol{\xi}, \dot{\boldsymbol{\xi}}, \ddot{\boldsymbol{\gamma}}, \ddot{\boldsymbol{\bar \gamma}}] \\
    &= -R^{a'}{}_{c'b'd'} \dot{\gamma}^{c'} \dot{\gamma}^{d'} \int_{\tau_0}^{\tau_1} \ud \tau_2 (\tau_1 - \tau_2) \pb{\gamma} H^{b'}{}_{b''} S^{b''} [\boldsymbol{\xi}, \dot{\boldsymbol{\xi}}, \ddot{\boldsymbol{\gamma}}, \ddot{\boldsymbol{\bar \gamma}}] + S^{a'} [\boldsymbol{\xi}, \dot{\boldsymbol{\xi}}, \ddot{\boldsymbol{\gamma}}, \ddot{\boldsymbol{\bar \gamma}}],
  \end{split}
\end{equation}
by the boundary conditions in Eq.~\eqref{eqn:K_H_bc} and by Eq.~\eqref{eqn:jacobi}.
Thus, we find that Eq.~\eqref{eqn:x_soln} gives a solution to Eq.~\eqref{eqn:ddot_xi} for arbitrary initial conditions.

We have therefore computed the solutions to the generalization of the geodesic deviation equation (where there are acceleration terms) to second order in the separation and its derivative.
This allows us to derive the explicit form of the curve deviation persistent observable, which is defined as

\begin{equation}
  \Delta \xi^{a'}_{\textrm{CD}} \equiv \xi^{a'} - \pb{\gamma} g^{a'}{}_a [\xi^a + (\tau_1 - \tau_0) \dot{\xi}^a] - \int_{\tau_0}^{\tau_1} \ud \tau_2 \int_{\tau_0}^{\tau_2} \ud \tau_3 \pb{\gamma} g^{a'}{}_{a'''} \left[g^{a'''}{}_{\bar a'''} \ddot{\bar \gamma}^{\bar a'''} - \ddot{\gamma}^{a'''}\right].
\end{equation}
Using the same notation as in Eq.~\eqref{eqn:curve_dev_vars}, we find that

\begin{subequations} \label{eqn:curve_dev_results}
  \begin{align}
    \Delta K^{a'}{}_b &= \pb{\gamma} K^{a'}{}_b - \pb{\gamma} g^{a'}{}_b, \displaybreak[0] \\
    \Delta H^{a'}{}_b &= \pb{\gamma} H^{a'}{}_b - \pb{\gamma} g^{a'}{}_b, \displaybreak[0] \\
    L^{a'}{}_{bc} &= -\int_{\tau_0}^{\tau_1} \ud \tau_2 (\tau_1 - \tau_2) \pb{\gamma} H^{a'}{}_{a''} \pb{\gamma} K^{b''}{}_{(b|} \dot{\gamma}^{d''} \bigg(\nabla_{(e''} R^{a''}{}_{c'')b''d''} \dot{\gamma}^{e''} \pb{\gamma} K^{c''}{}_{|c)} + 2 R^{a''}{}_{c''b''d''} \frac{\uD}{\ud \tau_2} \pb{\gamma} K^{c''}{}_{|c)}\bigg), \displaybreak[0] \\
    N^{a'}{}_{bc} &= -\int_{\tau_0}^{\tau_1} \ud \tau_2 (\tau_1 - \tau_2) \pb{\gamma} H^{a'}{}_{a''} \dot{\gamma}^{d''} \nonumber \\
    &\hspace{4.5em}\times \Bigg\{(\tau_2 - \tau_0) \left(\nabla_{(e''} R^{a''}{}_{c'')b''d''} + \nabla_{(e''} R^{a''}{}_{b'')c''d''}\right) \dot{\gamma}^{e''} \pb{\gamma} K^{b''}{}_b \pb{\gamma} H^{c''}{}_c \nonumber \\
    &\hspace{6.3em}+ 2 R^{a''}{}_{c''b''d''} \Bigg(\pb{\gamma} K^{b''}{}_b \frac{\uD}{\ud \tau_2} \left[(\tau_2 - \tau_0) \pb{\gamma} H^{c''}{}_c\right] + (\tau_2 - \tau_0) \pb{\gamma} H^{b''}{}_c \frac{\uD}{\ud \tau_2} \pb{\gamma} K^{c''}{}_b\Bigg)\Bigg\}, \displaybreak[0] \\
    M^{a'}{}_{bc} &= -\int_{\tau_0}^{\tau_1} \ud \tau_2 (\tau_1 - \tau_2) (\tau_2 - \tau_0) \pb{\gamma} H^{a'}{}_{a''} \pb{\gamma} H^{b''}{}_{(b|} \dot{\gamma}^{d''} \Bigg\{(\tau_2 - \tau_0) \nabla_{(e''} R^{a''}{}_{c'')b''d''} \dot{\gamma}^{e''} \pb{\gamma} H^{c''}{}_{|c)} \nonumber \\
    &\hspace{21.7em}+ 2 R^{a''}{}_{c''b''d''} \frac{\uD}{\ud \tau_2} \left[(\tau_2 - \tau_0) \pb{\gamma} H^{c''}{}_{|c)}\right]\Bigg\}.
  \end{align}
\end{subequations}
\end{widetext}

Again, we note that we are making the assumption that the separation vector is defined using the isochronous correspondence introduced in Sec.~\ref{sec:narrow}.
This is in contrast to the ``normal'' correspondence, where the separation vector is always orthogonal to the 4-velocity of one of the worldlines.
As was noted in Sec.~\ref{sec:traditional}, in the normal correspondence, there can be a difference between the final proper times along the worldlines (an effect which is absent in the isochronous correspondence).
However, using the isochronous correspondence, as we have done throughout most of this paper, does not mean that we have lost any information: here, the difference in proper time is encoded instead in the nonorthogonality of the final separation.

\subsection{Holonomies}

We now consider the computation of the holonomy of transport of linear and angular momentum using Eq.~\eqref{eqn:kappa_transport}.
As in Eq.~\eqref{eqn:kappa_transport_bundle}, we denote by $\vkappa{\nabla}_a$ the connection on the angular momentum bundle for arbitrary $\varkappa$.

First, we calculate $\vkappa{R}{}^A{}_{Bcd}$.
Note that any tensor $Z_{abcd} = V_{a[c} W_{d]b}$, where $V_{ab}$ and $W_{ab}$ are symmetric, satisfies

\begin{equation}
  Z_{abcd} = \frac{1}{2} (V_{ac} W_{db} - V_{ad} W_{bc}) = \frac{1}{2} (V_{ac} W_{db} - V_{da} W_{cb}).
\end{equation}
Both expressions on the right-hand side of this equation are differences of cyclic permutations of $bcd$ and $acd$, respectively, so they vanish under cyclic permutation:

\begin{align}
  Z_{a[bcd]} &= \frac{1}{3} (Z_{abcd} + Z_{acdb} + Z_{adbc}) = 0, \\
  Z_{[a|b|cd]} &= \frac{1}{3} (Z_{abcd} + Z_{cbda} + Z_{dbac}) = 0.
\end{align}
Thus, we find that $\vkappa{K}_{abcd}$ satisfies the first Bianchi identity:

\begin{equation} \label{eqn:K_bianchi}
  \vkappa{K}_{a[bcd]} = \vkappa{K}_{[a|b|cd]} = 0
\end{equation}
(note that this is true even though $\vkappa{K}_{abcd} \not\propto \vkappa{K}_{bacd}$).
This gives

\begin{equation}
  2 \vkappa{K}_{a[cd]b} = \vkappa{K}_{acdb} + \vkappa{K}_{adbc} = -\vkappa{K}_{abcd},
\end{equation}
and so we find

\begin{widetext}

\begin{equation} \label{eqn:mathfrak_kappa_R}
  \vkappa{R}{}^A{}_{Cef} = \begin{pmatrix}
    R^a{}_{cef} - 2 \vkappa{K}{}^a{}_{cef} & 2 \nabla_{[e} \vkappa{K}{}^a{}_{f]cd} \\
    0 & 2 \delta^{[a}{}_{[c} R^{b]}{}_{d]ef} + 4 \delta^{[a}{}_{[e} \vkappa{K}{}^{b]}{}_{f]cd}
  \end{pmatrix}.
\end{equation}

Given the parallel propagators with respect to $\vkappa{\nabla}_a$, the generic holonomy for any value of $\varkappa$ will be

\begin{equation} \label{eqn:kappa_holonomy}
  \vkappa{\Lambda}{}^A{}_C (\gamma, \bar{\gamma}; \tau_1) = \begin{pmatrix}
    \delta^a{}_c & 0 \\
    0 & \delta^{[a}{}_d \delta^{b]}{}_d
  \end{pmatrix} + \int_{\tau_0}^{\tau_1} \ud \tau_2 \xi^{[e''} \dot{\gamma}^{f'']} \begin{pmatrix}
    \smallunderset{PP}{\vkappa \Omega}{}^a{}_{ce''f''} (\gamma) & \smallunderset{PJ}{\vkappa \Omega}{}^a{}_{cde''f''} (\gamma) \vspace{0.5em} \\
    \smallunderset{JP}{\vkappa \Omega}{}^{ab}{}_{ce''f''} (\gamma) & \smallunderset{JJ}{\vkappa \Omega}{}^{ab}{}_{cde''f''} (\gamma)
  \end{pmatrix} + O(\boldsymbol{\xi}, \dot{\boldsymbol \xi})^2,
\end{equation}
where we have used Eq.~\eqref{eqn:holo_narrow_soln} to arrive at Eq.~\eqref{eqn:kappa_holonomy}, and we have defined

\begin{subequations} \label{eqn:kappa_hol_Omegas}
  \begin{align}
    \smallunderset{PP}{\vkappa \Omega}{}^a{}_{ce'f'} (\gamma) &= \pb{\gamma} \smallunderset{PP}{\vkappa{g}}{}^a{}_{a'} \left[\left(R^{a'}{}_{c'e'f'} - 2 \vkappa{K}{}^{a'}{}_{c'e'f'}\right) \pb{\gamma} \smallunderset{PP}{\vkappa g}{}^{c'}{}_c + 2 \nabla_{e'} \vkappa{K}{}^{a'}{}_{f'c'd'} \pb{\gamma} \smallunderset{JP}{\vkappa g}{}^{c'd'}{}_c\right] \nonumber \\
    &\hspace{1em}+ 2 \pb{\gamma} \smallunderset{PJ}{\vkappa{g}}{}^a{}_{a'b'} \left[\delta^{a'}{}_{c'} R^{b'}{}_{d'e'f'} + 2 \delta^{a'}{}_{e'} \vkappa{K}{}^{b'}{}_{f'c'd'}\right] \pb{\gamma} \smallunderset{JP}{\vkappa g}{}^{c'd'}{}_c, \displaybreak[0] \\
    \smallunderset{PJ}{\vkappa \Omega}{}^a{}_{cde'f'} (\gamma) &= \pb{\gamma} \smallunderset{PP}{\vkappa{g}}{}^a{}_{a'} \left[\left(R^{a'}{}_{c'e'f'} - 2 \vkappa{K}{}^{a'}{}_{c'e'f'}\right) \pb{\gamma} \smallunderset{PJ}{\vkappa g}{}^{c'}{}_{cd} + 2 \nabla_{e'} \vkappa{K}{}^{a'}{}_{f'c'd'} \pb{\gamma} \smallunderset{JJ}{\vkappa g}{}^{c'd'}{}_{cd}\right] \nonumber \\
    &\hspace{1em}+ 2 \pb{\gamma} \smallunderset{PJ}{\vkappa{g}}{}^a{}_{a'b'} \left[\delta^{a'}{}_{c'} R^{b'}{}_{d'e'f'} + 2 \delta^{a'}{}_{e'} \vkappa{K}{}^{b'}{}_{f'c'd'}\right] \pb{\gamma} \smallunderset{JJ}{\vkappa g}{}^{c'd'}{}_{cd}, \displaybreak[0] \\
    \smallunderset{JP}{\vkappa \Omega}{}^{ab}{}_{ce'f'} (\gamma) &= \pb{\gamma} \smallunderset{JP}{\vkappa{g}}{}^{ab}{}_{a'} \left[\left(R^{a'}{}_{c'e'f'} - 2 \vkappa{K}{}^{a'}{}_{c'e'f'}\right) \pb{\gamma} \smallunderset{PP}{\vkappa g}{}^{c'}{}_c + 2 \nabla_{e'} \vkappa{K}{}^{a'}{}_{f'c'd'} \pb{\gamma} \smallunderset{JP}{\vkappa g}{}^{c'd'}{}_c\right] \nonumber \\
    &\hspace{1em}+ 2 \pb{\gamma} \smallunderset{JJ}{\vkappa{g}}{}^{ab}{}_{a'b'} \left[\delta^{a'}{}_{c'} R^{b'}{}_{d'e'f'} + 2 \delta^{a'}{}_{e'} \vkappa{K}{}^{b'}{}_{f'c'd'}\right] \pb{\gamma} \smallunderset{JP}{\vkappa g}{}^{c'd'}{}_c, \displaybreak[0] \\
    \smallunderset{JJ}{\vkappa \Omega}{}^{ab}{}_{cde'f'} (\gamma) &= \pb{\gamma} \smallunderset{JP}{\vkappa{g}}{}^{ab}{}_{a'} \left[\left(R^{a'}{}_{c'e'f'} - 2 \vkappa{K}{}^{a'}{}_{c'e'f'}\right) \pb{\gamma} \smallunderset{PJ}{\vkappa g}{}^{c'}{}_{cd} + 2 \nabla_{e'} \vkappa{K}{}^{a'}{}_{f'c'd'} \pb{\gamma} \smallunderset{JJ}{\vkappa g}{}^{c'd'}{}_{cd}\right] \nonumber \\
    &\hspace{1em}+ 2 \pb{\gamma} \smallunderset{JJ}{\vkappa g}{}^{ab}{}_{a'b'} \Big[\delta^{a'}{}_{c'} R^{b'}{}_{d'e'f'} + 2 \delta^{a'}{}_{e'} \vkappa{K}{}^{b'}{}_{f'c'd'}\Big] \pb{\gamma} \smallunderset{JJ}{\vkappa g}{}^{c'd'}{}_{cd}.
  \end{align}
\end{subequations}
\end{widetext}
In most cases, we cannot analytically solve for these parallel propagators nonperturbatively in the Riemann tensor.
The results presented in Sec.~\ref{sec:memory_overview} are perturbative, assuming the curvature is weak along the worldline.
In such a case, solutions to Eq.~\eqref{eqn:U_int} can be truncated at a low order in the Riemann tensor.

\subsubsection{Affine transport holonomy} \label{sec:affine}

Now, we specialize to the case of affine transport.
We denote by $\mathring{\nabla}_a$ the connection on the linear and angular momentum bundle that is used for affine transport.
The parallel propagator with respect to this connection has an explicit solution, since the iterative solutions to Eqs.~\eqref{eqn:U_int} and~\eqref{eqn:U_inv_int} truncate by the nilpotence of $\pb{\gamma} \mathring{A}^A{}_{B'}$ to yield

\begin{subequations} \label{eqn:U_0}
  \begin{align}
    \pb{\gamma} \mathring{g}{}^{A'}{}_A &= \begin{pmatrix}
      \pb{\gamma} g^{a'}{}_a & 0 \\
      -2 \displaystyle\int_{\tau_0}^{\tau_1} \ud \tau_2 \pb{\gamma} g^{[a'}{}_{a''} \pb{\gamma} g^{b']}{}_a \dot{\gamma}^{a''} & \pb{\gamma} g^{[a'}{}_a \pb{\gamma} g^{b']}{}_b
    \end{pmatrix}, \displaybreak[0] \\
    \pb{\gamma} \mathring{g}{}^A{}_{A'} &= \begin{pmatrix}
      \pb{\gamma} g^a{}_{a'} & 0 \\
      2 \displaystyle\int_{\tau_0}^{\tau_1} \ud \tau_2 \pb{\gamma} g^{[a}{}_{a''} \pb{\gamma} g^{b]}{}_{a'} \dot{\gamma}^{a''} & \pb{\gamma} g^{[a}{}_{a'} \pb{\gamma} g^{b]}{}_{b'}
    \end{pmatrix}.
  \end{align}
\end{subequations}
Thus, we have that

\begin{widetext}
\begin{equation} \label{eqn:calR_0_mat}
  \pb{\gamma} \mathring{g}^A{}_{A'} \pb{\gamma} \mathring{g}^{C'}{}_C \mathring{R}^{A'}{}_{C'e'f'} = \begin{pmatrix}
    \pb{\gamma} g^a{}_{a'} \pb{\gamma} g^{c'}{}_c R^{a'}{}_{c'e'f'} & 0 \\
    2 \delta^{[a}{}_{c} \pb{\gamma} g^{b]}{}_{b'} R^{b'}{}_{g'e'f'} \displaystyle\int_{\tau_0}^{\tau_1} \ud \tau_2 \pb{\gamma} g^{g'}{}_{g''} \dot{\gamma}^{g''} & 2 \delta^{[a}{}_{[c} \pb{\gamma} g^{b]}{}_{|b'|} \pb{\gamma} g^{d'}{}_{d]} R^{b'}{}_{d'e'f'}
  \end{pmatrix}
\end{equation}
which yields, by Eq.~\eqref{eqn:holo_narrow_soln} and an integration by parts,

\begin{equation} \label{eqn:0_holonomy_mat}
  \mathring{\Lambda}{}^A{}_C (\gamma, \bar{\gamma}; \tau_1) = \begin{pmatrix}
    \Lambda^a{}_c (\gamma, \bar{\gamma}; \tau_1) & 0 \\
    2 \displaystyle\int_{\tau_0}^{\tau_1} \ud \tau_2 \displaystyle\int_{\tau_2}^{\tau_1} \ud \tau_3 \delta^{[a}{}_c \pb{\gamma} g^{b]}{}_{b'''} R^{b'''}{}_{g'''e'''f'''} \pb{\gamma} g^{g'''}{}_{g''} \dot{\gamma}^{g''} \dot{\gamma}^{e'''} \xi^{f'''} & 2 \delta^{[a}{}_{[c} \Lambda^{b]}{}_{d]} (\gamma, \bar{\gamma}; \tau_1)
  \end{pmatrix} + O(\boldsymbol{\xi}, \dot{\boldsymbol \xi})^2.
\end{equation}
Now, we also have that

\begin{equation}
  \dot{\gamma}^{a''} = \pb{\gamma} g^{a''}{}_{a'''} \dot{\gamma}^{a'''} - \int_{\tau_2}^{\tau_3} \ud \tau_4 \pb{\gamma} g^{a''}{}_{a''''} \ddot{\gamma}^{a''''},
\end{equation}
and so we find that Eq.~\eqref{eqn:0_holonomy_mat} becomes

\begin{subequations} \label{eqn:0_holonomy_final}
  \begin{align}
    \smallunderset{PP}{\mathring \Lambda}{}^a{}_c (\gamma, \bar{\gamma}; \tau_1) &= \Lambda^a{}_c (\gamma, \bar{\gamma}; \tau_1) + O(\boldsymbol{\xi}, \dot{\boldsymbol \xi})^2, \\
    \smallunderset{JP}{\mathring \Lambda}{}^{ab}{}_c (\gamma, \bar{\gamma}; \tau_1) &= 2 \delta^{[a}{}_c \delta^{b]}{}_e \Bigg\{\pb{\gamma} g^e{}_{e'} \left[(\tau_1 - \tau_0) \dot{\xi}^{e'} - \xi^{e'}\right] + \xi^e \nonumber \\
                                                                                &\hspace{5.3em}- \int_{\tau_0}^{\tau_1} \ud \tau_2 \int_{\tau_2}^{\tau_1} \ud \tau_3 \pb{\gamma} g^e{}_{e'''} \Bigg(g^{e'''}{}_{\bar e'''} \ddot{\bar \gamma}^{\bar e'''} - \ddot{\gamma}^{e'''} \nonumber \\
                                                                                 &\hspace{17em}+ \ddot{\gamma}^{f'''} \int_{\tau_3}^{\tau_1} \ud \tau_4 \pb{\gamma} g^{e'''}{}_{e''''} \pb{\gamma} g^{f''''}{}_{f'''} R^{e''''}{}_{f''''g''''h''''} \dot{\gamma}^{g''''} \xi^{h''''}\Bigg)\Bigg\} \nonumber \\
                                                                                 &\hspace{1em}+ O(\boldsymbol{\xi}, \dot{\boldsymbol \xi})^2, \\
     \smallunderset{JJ}{\mathring \Lambda}{}^{ab}{}_{cd} (\gamma, \bar{\gamma}; \tau_1) &= 2 \delta^{[a}{}_{[c} \Lambda^{b]}{}_{d]} (\gamma, \bar{\gamma}; \tau_1) + O(\boldsymbol{\xi}, \dot{\boldsymbol \xi})^2.
  \end{align}
\end{subequations}
\end{widetext}
Note that the first two terms in the expression for $\smallunderset{JP}{\mathring \Lambda}{}^{ab}{}_c (\gamma, \bar{\gamma}; \tau_1)$ are related to the displacement memory observable, as they are written in terms of $\xi^{a'}$.
The remaining terms measure the acceleration of the worldlines and additional time integrals of the Riemann tensor.
Both $\smallunderset{PP}{\mathring \Lambda}{}^a{}_c (\gamma, \bar{\gamma}; \tau_1)$ and $\smallunderset{JJ}{\mathring \Lambda}{}^{ab}{}_{cd} (\gamma, \bar{\gamma}; \tau_1)$ depend upon just the usual holonomy, and therefore, they contain the same information as the Lorentz transformation observable.

\subsubsection{Dual Killing transport holonomy} \label{sec:dual_killing}

The holonomy for dual Killing transport similarly has a nonperturbative solution, because the parallel propagators with respect to this connection are related to the Jacobi propagators (assuming that $\gamma$ is geodesic).
To see how, suppose that we have some $\xi_a$ and $F_{ab}$ defined as tensor fields along $\gamma$ such that

\begin{equation} \label{eqn:Y_killing}
  Y_A \equiv \begin{pmatrix}
    \xi_a & F_{ab}
  \end{pmatrix}, \qquad \dot{\gamma}^b \half{\nabla}_b Y_A = 0,
\end{equation}
where $\half{\nabla}_a$ is the connection associated with dual Killing transport.
This implies $\xi_a$ and $F_{ab}$ satisfy

\begin{subequations}
  \begin{align}
    \dot{\gamma}^b \nabla_b \xi_a &= 2 \dot{\gamma}^b F_{ba}, \label{eqn:D_xi} \\
    \dot{\gamma}^c \nabla_c F_{ab} &= \frac{1}{2} R^d{}_{cab} \xi_d \dot{\gamma}^c. \label{eqn:D_F}
  \end{align}
\end{subequations}
Then we have that (as $\gamma$ is geodesic)

\begin{equation} \label{eqn:D2_xi}
  (\dot{\gamma}^c \nabla_c) (\dot{\gamma}^d \nabla_d) \xi_a = -R^b{}_{cad} \dot{\gamma}^c \dot{\gamma}^d \xi_b.
\end{equation}
Note that by raising $a$ (which commutes with $\dot{\gamma}^b \nabla_b$), we obtain the linearized version of the geodesic deviation equation.
The Jacobi propagators therefore give the solution to Eq.~\eqref{eqn:D2_xi}.
By using~\eqref{eqn:jacobi}, we have

\begin{equation} \label{eqn:xi_soln}
  \xi_{a'} = \pb{\gamma} K_{a'}{}^a \xi_a + 2 (\tau_1 - \tau_0) \dot{\gamma}^a \pb{\gamma} H_{a'}{}^b F_{ab}
\end{equation}
(this follows from the fact that $g^a{}_{a'} = g_{a'b'} g^{ab} g^{b'}{}_b$, again a consequence of the compatibility of the metric and the Levi-Civita connection).
Integrating Eq.~\eqref{eqn:D_F}, we find that

\begin{widetext}

\begin{equation} \label{eqn:F_soln}
  F_{a'b'} = \pb{\gamma} g^a{}_{a'} \pb{\gamma} g^b{}_{b'} F_{ab} + \frac{1}{2} \int_{\tau_0}^{\tau_1} \ud \tau_2 \pb{\gamma} g^{a''}{}_{a'} \pb{\gamma} g^{b''}{}_{b'} R^{c''}{}_{d''a''b''} \dot{\gamma}^{d''} \xi_{c''}.
\end{equation}
Equations~\eqref{eqn:xi_soln} and~\eqref{eqn:F_soln} give $\xi_{a'}$ and $F_{a'b'}$ as linear functions of $\xi_a$ and $F_{ab}$, and we can use them to write the parallel propagator as follows:

\begin{equation} \label{eqn:U_half_inv}
  \pb{\gamma} \half{g}{}^A{}_{A'} = \begin{pmatrix}
    \pb{\gamma} K_{a'}{}^a & \dfrac{1}{2} \displaystyle\int_{\tau_0}^{\tau_1} \ud \tau_2 \pb{\gamma} K_{c''}{}^a R^{c''}{}_{d''a''b''} \dot{\gamma}^{d''} \pb{\gamma} g^{a''}{}_{a'} \pb{\gamma} g^{b''}{}_{b'} \\[0.8em]
    2 (\tau_1 - \tau_0) \dot{\gamma}^{[a} \pb{\gamma} H_{a'}{}^{b]} & \pb{\gamma} g^{[a}{}_{a'} \pb{\gamma} g^{b]}{}_{b'} + \displaystyle\int_{\tau_0}^{\tau_1} \ud \tau_2 (\tau_2 - \tau_0) \dot{\gamma}^{[a} \pb{\gamma} H_{c''}{}^{b]} R^{c''}{}_{d''a''b''} \dot{\gamma}^{d''} \pb{\gamma} g^{a''}{}_{a'} \pb{\gamma} g^{b''}{}_{b'}
  \end{pmatrix}.
\end{equation}
It is possible to invert this matrix, but a simpler approach is to switch $\tau_0$ with $\tau_1$, which yields

\begin{equation} \label{eqn:U_half}
  \pb{\gamma} \half{g}{}^{A'}{}_A = \begin{pmatrix}
    \pb{\gamma} K_a{}^{a'} & -\dfrac{1}{2} \displaystyle\int_{\tau_0}^{\tau_1} \ud \tau_2 \pb{\gamma} K_{c''}{}^{a'} R^{c''}{}_{d''a''b''} \dot{\gamma}^{d''} \pb{\gamma} g^{a''}{}_a \pb{\gamma} g^{b''}{}_b \\[0.8em]
    -2 (\tau_1 - \tau_0) \dot{\gamma}^{[a'} \pb{\gamma} H_a{}^{b']} & \pb{\gamma} g^{[a'}{}_a \pb{\gamma} g^{b']}{}_b + \displaystyle \int_{\tau_0}^{\tau_1} \ud \tau_2 (\tau_1 - \tau_2) \dot{\gamma}^{[a'} \pb{\gamma} H_{c''}{}^{b']} R^{c''}{}_{d''a''b''} \dot{\gamma}^{d''} \pb{\gamma} g^{a''}{}_a \pb{\gamma} g^{b''}{}_b
  \end{pmatrix}.
\end{equation}
Note that, to zeroth order in the Riemann tensor, these two equations agree with Eq.~\eqref{eqn:U_0}.

To complete the calculation of the holonomy for dual Killing transport, we further simplify our expression for $\half{R}{}^A{}_{Bcd}$.
Note that $\smallunderset{PP}{\half R}{}^a{}_{cef} = 0$, and

\begin{equation}
  2 \nabla_{[e} R_{|a|f]cd} = \nabla_e R_{afcd} + \nabla_f R_{eacd} = \nabla_a R_{efcd},
\end{equation}
by the second Bianchi identity, so $\smallunderset{PJ}{\half R}{}^a{}_{cdef} = \frac{1}{2} \nabla^a R_{efcd} = \frac{1}{2} \nabla^a R_{cdef}$.
Using the same notation as in Eq.~\eqref{eqn:kappa_holonomy}, we obtain our final result in terms of the parallel and Jacobi propagators and the curvature along the worldline:

\begin{subequations} \label{eqn:half_Lambda}
  \begin{align}
    \smallunderset{PP}{\half \Omega}{}^a{}_{ce'f'} (\gamma) &= -2 (\tau_1 - \tau_0) \Bigg[\frac{1}{4} \pb{\gamma} K_{a'}{}^a \nabla^{a'} R_{e'f'c'd'} + \int_{\tau_0}^{\tau_1} \ud \tau_2 \pb{\gamma} K_{g''}{}^a R^{g''}{}_{h''a''b''} \dot{\gamma}^{h''} \pb{\gamma} g^{b''}{}_{b'} \pb{\gamma} g^{a''}{}_{[c'} R^{b'}{}_{d']e'f'} \nonumber \\
    &\hspace{6.7em}+ ([c'd'] \leftrightarrow [e'f'])\Bigg] \dot{\gamma}^{c'} \pb{\gamma} H_c{}^{d'}, \displaybreak[0] \\
    \smallunderset{PJ}{\half \Omega}{}^a{}_{cde'f'} (\gamma) &= \Bigg[\frac{1}{4} \pb{\gamma} K_{a'}{}^a \nabla^{a'} R_{e'f'c'd'} + \int_{\tau_0}^{\tau_1} \ud \tau_2 \pb{\gamma} K_{g''}{}^a R^{g''}{}_{h''a''b''} \dot{\gamma}^{h''} \pb{\gamma} g^{b''}{}_{b'} \pb{\gamma} g^{a''}{}_{[c'} R^{b'}{}_{d']e'f'} \nonumber \\
                                                                             &\hspace{1.7em}+ ([c'd'] \leftrightarrow [e'f'])\Bigg] \Bigg[\pb{\gamma} g^{c'}{}_c \pb{\gamma} g^{d'}{}_d + \int_{\tau_0}^{\tau_1} \ud \tau_2 (\tau_1 - \tau_2) \dot{\gamma}^{c'} \pb{\gamma} H_{g''}{}^{d'} R^{g''}{}_{h''c''d''} \dot{\gamma}^{h''} \pb{\gamma} g^{c''}{}_c \pb{\gamma} g^{d''}{}_d\Bigg], \\
    \smallunderset{JP}{\half \Omega}{}^{ab}{}_{ce'f'} (\gamma) &= -2 (\tau_1 - \tau_0) \Bigg\{\frac{1}{2} (\tau_1 - \tau_0) \dot{\gamma}^{[a} \pb{\gamma} H_{a'}{}^{b]} \nabla^{a'} R_{e'f'c'd'} \nonumber \\
                                                                             &\hspace{6.7em}+ 2 \Bigg[\pb{\gamma} g^{[a}{}_{a'} \pb{\gamma} g^{b]}{}_{b'} + \int_{\tau_0}^{\tau_1} \ud \tau_2 (\tau_2 - \tau_0) \dot{\gamma}^{[a} \pb{\gamma} H_{g''}{}^{b]} R^{g''}{}_{h''a''b''} \dot{\gamma}^{h''} \pb{\gamma} g^{a''}{}_{a'} \pb{\gamma} g^{b''}{}_{b'}\Bigg] \delta^{a'}{}_{[c'} R^{b'}{}_{d']e'f'} \nonumber \\
    &\hspace{6.7em}+ ([c'd'] \leftrightarrow [e'f'])\Bigg\} \dot{\gamma}^{c'} \pb{\gamma} H_c{}^{d'}, \displaybreak[0] \\
    \smallunderset{JJ}{\half \Omega}{}^{ab}{}_{cde'f'} (\gamma) &= \Bigg\{\frac{1}{2} (\tau_1 - \tau_0) \dot{\gamma}^{[a} \pb{\gamma} H_{a'}{}^{b]} \nabla^{a'} R_{e'f'c'd'} \nonumber \\
                                                                             &\hspace{1.86em}+ 2 \Bigg[\pb{\gamma} g^{[a}{}_{a'} \pb{\gamma} g^{b]}{}_{b'} + \int_{\tau_0}^{\tau_1} \ud \tau_2 (\tau_2 - \tau_0) \dot{\gamma}^{[a} \pb{\gamma} H_{g''}{}^{b]} R^{g''}{}_{h''a''b''} \dot{\gamma}^{h''} \pb{\gamma} g^{a''}{}_{a'} \pb{\gamma} g^{b''}{}_{b'}\Bigg] \delta^{a'}{}_{[c'} R^{b'}{}_{d']e'f'} \nonumber \\
                                                                             &\hspace{1.86em}+ ([c'd'] \leftrightarrow [e'f'])\Bigg\} \Bigg[\pb{\gamma} g^{c'}{}_c \pb{\gamma} g^{d'}{}_d + \int_{\tau_0}^{\tau_1} \ud \tau_2 (\tau_1 - \tau_2) \dot{\gamma}^{c'} \pb{\gamma} H_{g''}{}^{d'} R^{g''}{}_{h''c''d''} \dot{\gamma}^{h''} \pb{\gamma} g^{c''}{}_c \pb{\gamma} g^{d''}{}_d\Bigg].
  \end{align}
\end{subequations}
Here ``$+ ([c'd'] \leftrightarrow [e'f'])$'' means ``add all the previous terms in the sum, but with $[c'd']$ and $[e'f']$ switched.''

This concludes our calculation of the holonomy observables.
In a subsequent paper in this series, we will be considering this expression in the nonperturbative regime in spacetimes where the Jacobi propagators are known, in particular in plane wave spacetimes.

\end{widetext}

\subsection{Spinning particles} \label{sec:spin_comp}

We now consider the procedure outlined in Sec.~\ref{sec:spinning_observables}: an observer measures the separation $\xi^a$ from an initially comoving spinning test particle, as well as its linear momentum $p^a$ and spin per unit mass $s^a$.
At some later point in time, the observer performs these measurements again.
The persistent observables in this case are the differences between the initial and final measurements.

In order to compute these observables, we need to determine the worldline of the spinning particle.
First, note that the Mathisson-Papapetrou equations [Eq.~\eqref{eqn:mp}] do not form a fully determined system of equations, as they contain 13 variables (four in $p^{\bar a}$, six in $j^{\bar a \bar b}$, and three in $\dot{\bar \gamma}^{\bar a}$), but only 10 equations.
To solve for all of these variables (in particular $\dot{\bar \gamma}^{\bar a}$), we would need three more equations, which are given by so-called \emph{spin-supplementary conditions}.
A commonly used spin supplementary condition is the Tulczyjew condition~\cite{Tulczyjew1959}, which is given by enforcing

\begin{equation} \label{eqn:tulczyjew}
  j^{\bar a \bar b} p_{\bar b} = 0
\end{equation}
along the worldline, for all $\tau$.
This says that the mass dipole moment, measured in the rest frame determined by $p^a$, is zero.
This lends itself to a convenient definition of intrinsic spin per unit mass:
\begin{equation} \label{eqn:s_def}
  s^{\bar a} \equiv -\frac{1}{2 p^{\bar e} p_{\bar e}} \epsilon^{\bar a \bar b \bar c \bar d} p_{\bar b} j_{\bar c \bar d}.
\end{equation}

At this point, we refer the reader to a derivation of the acceleration of the spinning particle that we give in Appendix~\ref{app:mp_solution}, and merely present the results here.
First, the intrinsic spin per unit mass is merely parallel transported along $\bar{\gamma}$ to order spin squared:

\begin{equation}
  s^{\bar a'} = \pb{\bar \gamma} g^{\bar a'}{}_{\bar a} s^{\bar a} + O(\boldsymbol{s})^2.
\end{equation}
Moreover, at all times $\tau$ along the worldline, the momentum $p^{\bar a}$ is related to the 4-velocity $\dot{\bar \gamma}^{\bar a}$ by

\begin{equation}
  p^{\bar a} = m \dot{\bar \gamma}^{\bar a} + O(\boldsymbol{s})^2,
\end{equation}
where the mass $m$ is constant (again to order spin squared).
Finally, the acceleration of the spinning particle is given by

\begin{equation}
  \ddot{\bar \gamma}^{\bar a} = -(R^*)^{\bar a}{}_{\bar c \bar b \bar d} \dot{\bar \gamma}^{\bar c} \dot{\bar \gamma}^{\bar d} s^{\bar b} + O(\boldsymbol{s})^2.
\end{equation}

At this point we can compute the observables discussed in Sec.~\ref{sec:spinning_observables}.
These are given in terms of the initially measured momentum $p^a$ and intrinsic spin $s^a$, which are given by

\begin{equation}
  p^a = m g^a{}_{\bar a} \dot{\bar \gamma}^{\bar a} = m \dot{\gamma}^a, \qquad s^a = g^a{}_{\bar a} s^{\bar a},
\end{equation}
assuming the observer and spinning particle are initially comoving.
For the intrinsic spin, we have that

\begin{equation}
  \begin{split}
    \Delta s^{a'} &= (g^{a'}{}_{\bar a'} \pb{\bar \gamma} g^{\bar a'}{}_{\bar a} g^{\bar a}{}_a - \pb{\gamma} g^{a'}{}_a) s^a + O(\boldsymbol{s})^2 \\
    &= \pb{\gamma} g^{a'}{}_a \left[\left(\Lambda^{-1}\right){}^a{}_b (\gamma, \bar{\gamma}; \tau_1) - \delta^a{}_b\right] s^b + O(\boldsymbol{s})^2, \\
  \end{split}
\end{equation}
because the spin is parallel transported throughout the procedure to measure this persistent observable.
Thus, a nonzero $\Delta s^a$ arises because of a nontrivial holonomy.

The separation evolves using the general curve deviation equation~\eqref{eqn:ddot_xi}.
Unlike in the case of the curve deviation observable, the observer does not compare the final separation with the predicted separation in flat space, but instead with the initial separation:

\begin{equation}
  \Delta \xi^{a'}_{\textrm S} \equiv \xi^{a'} - \pb{\gamma} g^{a'}{}_a \xi^a.
\end{equation}
Note that, for the momentum, Eq.~\eqref{eqn:utilde} implies that

\begin{equation} \label{eqn:Delta_p}
  \begin{split}
    \Delta p^{a'} &= m \left(\dot{\xi}^{a'} + \dot{\gamma}^{a'}\right) - \pb{\gamma} g^{a'}{}_a p^a + O(\boldsymbol{s})^2 \\
    &= m \dot{\xi}^{a'} + O(\boldsymbol{s})^2 = m \frac{\uD}{\ud \tau_1} \Delta \xi^{a'}_{\textrm S} + O(\boldsymbol{s})^2,
  \end{split}
\end{equation}
which proves Eq.~\eqref{eqn:Delta_p_decomp}; therefore, the computation of $\Delta p^{a'}$ is trivial once $\Delta \xi^{a'}_{\textrm{S}}$ is known.

To compute $\Delta \xi_{\textrm S}^{a'}$, we first need to calculate the acceleration of the spinning test particle to the relevant order:

\begin{widetext}
\begin{equation} \label{eqn:g_ddot_gamma_spin}
  \begin{split}
    g^{a'}{}_{\bar a'} \ddot{\bar \gamma}^{\bar a'} &= -g^{a'}{}_{\bar a'} (R^*)^{\bar a'}{}_{\bar c' \bar b' \bar d'} \dot{\bar \gamma}^{\bar c'} \dot{\bar \gamma}^{\bar d'} \pb{\bar \gamma} g^{\bar b'}{}_{\bar b} g^{\bar b}{}_b s^b + O(\boldsymbol{s})^2 \\
    &= -\left[(R^*)^{a'}{}_{c'b'd'} + \xi^{e'} \nabla_{e'} (R^*)^{a'}{}_{c'b'd'} + O(\boldsymbol{\xi}^2)\right] \left[\dot{\gamma}^{c'} \dot{\gamma}^{d'} + 2 \dot{\gamma}^{(c'} \dot{\xi}^{d')} + O(\boldsymbol{\xi}^2)\right] \pb{\gamma} g^{b'}{}_e (\Lambda^{-1})^e{}_b (\gamma, \bar{\gamma}; \tau_1) s^b + O(\boldsymbol{s})^2.
  \end{split}
\end{equation}
To derive Eq.~\eqref{eqn:g_ddot_gamma_spin}, we have used the definition of the holonomy, Eq.~\eqref{eqn:utilde}, and the coincidence limit of the Riemann tensor.
Now, we use Eq.~\eqref{eqn:holo_narrow_soln} and the solution to the geodesic equation in Eq.~\eqref{eqn:x_soln} to write Eq.~\eqref{eqn:g_ddot_gamma_spin} in terms of $\xi^a$ and $s^a$:

\begin{equation}
  \begin{split}
    g^{a'}{}_{\bar a'} \ddot{\bar \gamma}^{\bar a'} = -\Bigg\{&(R^*)^{a'}{}_{c'b'd'} \pb{\gamma} g^{b'}{}_b \dot{\gamma}^{c'} \\
    &+ \Bigg[\pb{\gamma} K^{c'}{}_c \nabla_{c'} (R^*)^{a'}{}_{d'b'e'} \dot{\gamma}^{e'} \pb{\gamma} g^{b'}{}_b \\
    &\hspace{2em} - (R^*)^{a'}{}_{(c'|e'|d')} \int_{\tau_0}^{\tau_1} \ud \tau_2 R^{e''}{}_{b''f''g''} \left(\pb{\gamma} g^{e'}{}_{e''} \pb{\gamma} g^{b''}{}_b \dot{\gamma}^{c'} + 2 \dot{\gamma}^{b''} \pb{\gamma} g^{c'}{}_{e''} \pb{\gamma} g^{e'}{}_b\right) \dot{\gamma}^{g''} \pb{\gamma} K^{f''}{}_c\Bigg] \xi^c \\
    &+ O(\boldsymbol{\xi}^2)\Bigg\} s^b \dot{\gamma}^{d'} + O(\boldsymbol{s})^2.
  \end{split}
\end{equation}
Using Eq.~\eqref{eqn:x_soln}, we find that our observables [using the notation in Eq.~\eqref{eqn:spin_obs_decomp}] are given by

\begin{subequations} \label{eqn:spin_obs_final}
  \begin{align}
    \Sigma^{a'}{}_{bc} &= -\int_{\tau_0}^{\tau_1} \ud \tau_2 \pb{\gamma} g^{a'}{}_{a''} R^{a''}{}_{b''c''d''} \pb{\gamma} g^{b''}{}_b \pb{\gamma} K^{c''}{}_c \dot{\gamma}^{d''}, \displaybreak[0] \\
    \Upsilon^{a'}{}_b &= -\int_{\tau_0}^{\tau_1} \ud \tau_2 (\tau_1 - \tau_2) \pb{\gamma} H^{a'}{}_{a''} (R^*)^{a''}{}_{c''b''d''} \dot{\gamma}^{c''} \dot{\gamma}^{d''} \pb{\gamma} g^{b''}{}_b, \displaybreak[0] \\
    \Psi^{a'}{}_{bc} &= -\int_{\tau_0}^{\tau_1} \ud \tau_2 (\tau_1 - \tau_2) \pb{\gamma} H^{a'}{}_{a''} \Bigg[\pb{\gamma} K^{c''}{}_c \nabla_{c''} (R^*)^{a''}{}_{d''b''e''} \dot{\gamma}^{e''} \pb{\gamma} g^{b''}{}_b \nonumber \\
    &\hspace{13.25em} - (R^*)^{a''}{}_{(c''|e''|d'')} \int_{\tau_0}^{\tau_2} \ud \tau_3 R^{e'''}{}_{b'''f'''g'''} \Big(\pb{\gamma} g^{e''}{}_{e'''} \pb{\gamma} g^{b'''}{}_b \dot{\gamma}^{c''} \nonumber \\
    &\hspace{30.5em}+ 2 \dot{\gamma}^{b'''} \pb{\gamma} g^{c''}{}_{e'''} \pb{\gamma} g^{e''}{}_b\Big) \dot{\gamma}^{g'''} \pb{\gamma} K^{f'''}{}_c\Bigg] \dot{\gamma}^{d''}.
  \end{align}
\end{subequations}
\end{widetext}
As these results are given in terms of the Riemann tensor and the fundamental bitensors (parallel and Jacobi propagators), they can be computed with relative ease in spacetimes in which these bitensors are known.

\section{Discussion} \label{sec:discussion}

In this paper, we have introduced quantities that we called persistent gravitational wave observables, which are effects that share with the gravitational wave memory effect the feature of persistence after a burst of gravitational waves, but which are not necessarily associated with symmetries and conserved quantities at boundaries of spacetime.
After reviewing many of the currently known persistent observables from the literature, we presented three new observables:

\begin{enumerate}
\item the difference between the separation of two accelerating curves from the result expected in flat space, which we called ``curve deviation,''
\item the path dependence (or ``holonomy'') for two different methods for relating linear and angular momentum at different points (one inspired by how linear and angular momentum transform under a change of origin in flat space, and the other by the relationship between linear and angular momentum and Killing vectors), and
\item the difference between the initial and final separation, 4-momentum, and spin of a spinning test particle that is initially comoving with some observer.
\end{enumerate}
These observables measure the effects of the gravitational waves in a context where the spacetime transitions from a flat region, to a burst of gravitational waves, and then to another flat region.

We then provided the machinery with which one can calculate these observables in an arbitrary spacetime (which included reviewing the very powerful technique of covariant bitensors for understanding how tensor fields evolve along curves).
Extending the results of~\cite{Vines2014a, Vines2014b}, we used these techniques to compute the holonomy with respect to an arbitrary connection around a variety of curves, as well as the evolution of the separation vector between two arbitrary worldlines.
We then used these holonomies and the separation vector to compute our final results, which are in Eqs.~\eqref{eqn:curve_dev_results} for curve deviation, Eqs.~\eqref{eqn:0_holonomy_final} and~\eqref{eqn:half_Lambda} for two different methods of relating angular momentum at different points, and Eq.~\eqref{eqn:spin_obs_final} for the observables from a spinning test particle.
Here, in order to make calculations tractable analytically, we made the simplifying assumption that the worldlines were close.

A strength of these results are that they are not specialized to a particular spacetime.
Our results are written in terms of the ``fundamental bitensors,'' which are solutions to the equations of parallel transport (the parallel propagators) and linear geodesic deviation (the Jacobi propagators).
These bitensors are known in a handful of spacetimes; we will use this fact in a future paper to derive more explicit expressions in exact, nonlinear plane-wave spacetimes, in which these bitensors are known~\cite{Harte2012a}.
In spacetimes where the geodesic equation has explicit solutions, these persistent observables can even be computed without assuming that the neighboring worldlines are close.

We also presented explicit expressions assuming that the curvature is small where these observables are being measured, so we may linearize in the spacetime curvature.
This provides a connection to previous memory observables, which are typically discussed in this regime.
These last results, where we linearize in the Riemann tensor, are important for discussing one possibility for measuring these persistent observables.
Our results were given only in terms of various integrals, or alternatively, moments, of the Riemann tensor (and its derivatives) with respect to proper time.
Moreover, in the limit where the gravitational waves are plane waves, these linearized results simplify even further, and they can be written entirely in terms of one, two, and three time integrals of the Riemann tensor, when there is no acceleration, and more time integrals, otherwise.
As gravitational wave detectors effectively measure the Riemann tensor along their worldlines, these integrals of the Riemann tensor are (in principle) measurable.
This would allow for our persistent observables to be measured indirectly.

Finally, a natural regime to study persistent gravitational wave observables is near future null infinity; of particular interest are their falloffs in $1/r$ near null infinity.
Here, the contexts that are relevant for studying persistent observables are spacetimes that possess two nonradiative regions that are separated by a radiative region.
As the two nonradiative regions are no longer flat, it is possible that the observables in this paper will also measure parts of the spacetime curvature not related to the gravitational waves, and so will not qualify as persistent gravitational wave observables in this context.
In a future paper, we will discuss the persistent gravitational wave observables that arise near null infinity.

\section*{Acknowledgments}

\'E.\'E.F.\ and A.M.G.\ acknowledge the support of NSF Grants No. PHY-1404105 and PHY-1707800 to Cornell.
D.A.N.\ acknowledges the support of the Netherlands Organization for Scientific Research through the NWO VIDI Grant No.\ 639.042.612-Nissanke and thanks Tanja Hinderer and Samaya Nissanke for useful discussions.

\appendix

\section{Dualization of Arbitrary Tensors} \label{sec:dual}

Following Penrose and Rindler~\cite{Penrose1987, Penrose1988} we define left and right duals of tensors acting on either the first or last two indices:

\begin{subequations} \label{eqn:star_def}
  \begin{align}
    ({}^* Z)_{abc_1 \cdots c_s} &\equiv \frac{1}{2} \epsilon_{abde} Z^{de}{}_{c_1 \cdots c_s}, \\
    (Z^*)_{a_1 \cdots a_s bc} &\equiv \frac{1}{2} Z_{a_1 \cdots a_s}{}^{de} \epsilon_{debc}.
  \end{align}
\end{subequations}
In addition to this standard definition, they define another type of dual, which acts on the first or last indices,

\begin{subequations} \label{eqn:dagger_def}
  \begin{align}
    ({}^\dagger Z)_{abcd_1 \cdots d_s} &\equiv \epsilon_{eabc} Z^e{}_{d_1 \cdots d_s}, \\
    (Z^\dagger)_{a_1 \cdots a_s bcd} &\equiv Z_{a_1 \cdots a_s} {}^e \epsilon_{ebcd},
  \end{align}
\end{subequations}
and a dual acting on the first or last three indices,

\begin{subequations} \label{eqn:ddagger_def}
  \begin{align}
    ({}^\ddagger Z)_{ab_1 \cdots b_s} \equiv \frac{1}{6} \epsilon_{cdea} Z^{cde}{}_{b_1 \cdots b_s}, \\
    (Z^\ddagger)_{a_1 \cdots a_s b} \equiv \frac{1}{6} Z_{a_1 \cdots a_s}{}^{cde} \epsilon_{cdeb}.
  \end{align}
\end{subequations}
With these definitions, we have that

\begin{align}
  {}^{**} Z_{abc_1 \cdots c_s} &= -Z_{[ab]c_1 \cdots c_s}, &\; Z^{**}_{a_1 \cdots a_s bc} &= -Z_{a_1 \cdots a_s [bc]}, \\
  {}^{\ddagger\dagger} Z_{ab_1 \cdots b_s} &= Z_{ab_1 \cdots b_s}, &\; Z^{\dagger\ddagger}_{a_1 \cdots a_s b} &= Z_{a_1 \cdots a_s b}, \\
  {}^{\dagger\ddagger} Z_{abcd_1 \cdots d_s} &= Z_{[abc]d_1 \cdots d_s}, &\; Z^{\ddagger\dagger}_{a_1 \cdots a_s bcd} &= Z_{a_1 \cdots a_s [bcd]}.
\end{align}
In four dimensions, these are the only useful definitions of duals of arbitrary tensors.

\section{Coincidence limits} \label{sec:coincidence}

In this appendix, we briefly review coincidence limits and give expressions for coincidence limits that we have used earlier in this paper.
The \emph{coincidence limit} of a bitensor $T_{A_1 \cdots A_r B_1' \cdots B_s'}$ (where capital letters denote arbitrary bundle indices) is given by

\begin{equation}
  \left[T_{A_1 \cdots A_r B_1' \cdots B_s'}\right]_{x' \to x} \equiv \lim_{x' \to x} \hat{g}^{B_1'}{}_{B_1}\!\! \cdots\! \hat{g}^{B_s'}{}_{B_s} T_{A_1 \cdots A_r B_1' \cdots B_s'},
\end{equation}
for some parallel propagator $\hat{g}^{A'}{}_A$; it is trivial to show that this is independent of the parallel propagator that is used, which is why there is no parallel propagator on the left-hand side.
By convention, the indices inside the coincidence limit that are associated with the point whose limit is being taken (in this case $x'$) are treated as if they were at the limiting point (in this case $x$) for expressions outside of the brackets.
We use this notation throughout, following the review article of Poisson~\cite{Poisson2004}; simple examples can be seen below in Eq.~\eqref{eqn:synge_coincidence}.

We now list the coincidence limits we have used in this paper.
A general procedure for computing these coincidence limits is outlined in~\cite{Poisson2004}.
These expressions can also be found in~\cite{Vines2014b}.
For Synge's world function, the relevant coincidence limits are

\begin{subequations} \label{eqn:synge_coincidence}
  \begin{gather}
    \delta^a{}_b = \left[\sigma^a{}_b\right]_{x' \to x} = -\left[\sigma^a{}_{b'}\right]_{x' \to x}, \displaybreak[0] \\
    0 = \left[\sigma^a{}_{bc'}\right]_{x' \to x} = \left[\sigma^a{}_{b'c'}\right]_{x' \to x}, \displaybreak[0] \\
    -\frac{2}{3} R^a{}_{(c|b|d)} = \left[\sigma^a{}_{bc'd'}\right]_{x' \to x} = 2 \left[\sigma^a{}_{b'(c'd')}\right]_{x' \to x},
  \end{gather}
\end{subequations}
while for the parallel propagator, they are

\begin{subequations} \label{eqn:prop_coincidence}
  \begin{gather}
    0 = \left[\hat{\nabla}_c \hat{g}^{A'}{}_B\right]_{x' \to x} = \left[\hat{\nabla}_{c'} \hat{g}^{A'}{}_B\right]_{x' \to x}, \\[.5em]
    \begin{aligned}
      \frac{1}{2} \hat{R}^A{}_{Bcd} &= \left[\hat{\nabla}_{c'} \hat{\nabla}_{d'} \hat{g}^{A'}{}_B\right]_{x' \to x} = \left[\nabla_{c'} \nabla_d g^{A'}{}_B\right]_{x' \to x} \\
      &= -\left[\hat{\nabla}_c \hat{\nabla}_{d'} \hat{g}^{A'}{}_B\right]_{x' \to x} = -\left[\hat{\nabla}_c \hat{\nabla}_d \hat{g}^{A'}{}_B\right]_{x' \to x},
    \end{aligned} \displaybreak[0] \\
    \begin{aligned}
      \frac{2}{3} \hat{\nabla}_{(c} \hat{R}^A{}_{|B|d)e} &= \left[\nabla_{c'} \nabla_{d'} \nabla_{e'} \hat{g}^{A'}{}_B\right]_{x' \to x} \\
      &= 2 \left[\nabla_{(c'} \nabla_{d')} \nabla_e \hat{g}^{A'}{}_B\right]_{x' \to x}.
    \end{aligned}
  \end{gather}
\end{subequations}
Moreover, for any bitensor $T_{a_1 \cdots a_r b_1' \cdots b_s'}$~\cite{Vines2014b},

\begin{equation} \label{eqn:switch_coincidence}
  \left[T_{a_1 \cdots a_r b_1' \cdots b_s'}\right]_{x' \to x} = \left[T_{a_1' \cdots a_r' b_1 \cdots b_s}\right]_{x' \to x}.
\end{equation}
All of the coincidence limits which are needed in this paper can be derived from using this property of coincidence limits, Eq.~\eqref{eqn:synge_coincidence}, and Eq.~\eqref{eqn:prop_coincidence}.

\section{Solution to the Mathisson-Papapetrou equations} \label{app:mp_solution}

In this section, we review the solution to the Mathisson-Papapetrou equations, to linear order in the spin, by adapting a proof from~\cite{Semerak1999}.
Throughout this derivation, for convenience, we denote the worldline of the spinning particle by $\Gamma$, and use unadorned indices at $\Gamma(\tau)$ (where $\tau$ is arbitrary).

To begin, define the following notions of mass and mass ratio,

\begin{gather}
  \mathcal{M} (\tau) \equiv -p^a \dot{\Gamma}_a, \qquad m(\tau) \equiv \sqrt{-p^a p_a}, \\
  \mu(\tau) \equiv \mathcal{M}(\tau)/m(\tau),
\end{gather}
and the ``dynamical'' 4-velocity

\begin{equation}
  \mathcal{U}^a \equiv p^a/m(\tau).
\end{equation}
The definition of intrinsic spin per unit mass in Eq.~\eqref{eqn:s_def} obeys the equation

\begin{equation} \label{eqn:s_p}
  s^a p_a = 0.
\end{equation}
By the Tulczyjew condition, $s^a$ and $p^a$ are equivalent to $j^{ab}$, as

\begin{equation} \label{eqn:j_s}
  \begin{split}
    \epsilon^{abcd} p_c s_d &= -\frac{1}{2} \epsilon^{dabc} \mathcal{U}_c \epsilon_{defg} \mathcal{U}^e j^{fg} \\
    &= 3 \mathcal{U}^{[a} j^{bc]} \mathcal{U}_c \\
    &= -j^{ab}.
  \end{split}
\end{equation}
Thus, we have that Eq.~\eqref{eqn:mp} can be rewritten as

\begin{subequations}
  \begin{align}
    \dot{\mathcal U}^a &= -[\dot{m} (\tau)/m (\tau)] \mathcal{U}^a + (R^*)^a{}_{bcd} \dot{\Gamma}^b \mathcal{U}^c s^d, \label{eqn:mp_Udot} \\
    \dot{s}^a &= -[\dot{m} (\tau)/m (\tau)] s^a + \mathcal{U}^a (R^*)_{bcde} s^b \dot{\Gamma}^c \mathcal{U}^d s^e,
  \end{align}
\end{subequations}
where we have used Eq.~\eqref{eqn:j_s} and the orthogonality of $\mathcal{U}^a$ and $\dot{\mathcal{U}}^a$.
On contracting the first equation with $p_a$, we obtain

\begin{equation} \label{eqn:Mdot}
  \dot{m} (\tau) = -(R^*)_{abcd} p^a \dot{\Gamma}^b \mathcal{U}^c s^d,
\end{equation}
so the second equation reads

\begin{equation}
  \dot{s}^a = 2 s^{(a} (R^*)_{bcde} \mathcal{U}^{b)} \dot{\Gamma}^c \mathcal{U}^d s^e.
\end{equation}

Now, we suppose that $s^a$ is initially small, so we can linearize in it.
Because its derivative is also small, in fact $O(\boldsymbol{s}^2)$, we can linearize in $s^a$ along the entire worldline of the particle.
Using Eq.~\eqref{eqn:mp_Jdot}, as well as the derivative of the Tulczyjew condition, we have that

\begin{equation} \label{eqn:mp_gammadot}
  \begin{split}
    \dot{\Gamma}^a &= \mu(\tau) \mathcal{U}^a + \epsilon^{abcd} \dot{p}_b \mathcal{U}_c s_d \\
    &= \mu(\tau) \mathcal{U}^a + O(\boldsymbol{s}^2) \\
    &= \mathcal{U}^a + O(\boldsymbol{s}^2),
  \end{split}
\end{equation}
where in the second line we have used Eq.~\eqref{eqn:mp_Pdot} and in the third line we have used the fact that $\mu(\tau)$ is set by normalizing $\dot{\Gamma}^a \dot{\Gamma}_a = \mathcal{U}^a \mathcal{U}_a = -1$.
From Eq.~\eqref{eqn:Mdot}, we therefore have that

\begin{equation} \label{eqn:Mdot_pert}
  \begin{split}
    \dot{m} (\tau) &= -m (\tau) (R^*)_{abcd} \mathcal{U}^a \mathcal{U}^b \mathcal{U}^c s^d + O(\boldsymbol{s}^2) \\
    &= O(\boldsymbol{s}^2).
  \end{split}
\end{equation}
Putting together Eqs.~\eqref{eqn:mp_Udot},~\eqref{eqn:mp_gammadot}, and~\eqref{eqn:Mdot_pert}, we find that the acceleration of the spinning particle is given by

\begin{equation} \label{eqn:acc_mp}
  \ddot{\Gamma}^a = -(R^*)^a{}_{cbd} \dot{\Gamma}^c \dot{\Gamma}^d s^b + O(\boldsymbol{s}^2).
\end{equation}

\section{Algebraic Decomposition of Holonomies} \label{sec:algebraic}

In this section, we present a method of reducing the holonomy observable in Sec.~\ref{sec:holonomy} into more manageable pieces.
Our method is purely algebraic and applies to general matrices on the linear and angular momentum bundle.
Consider first any matrix $A^A{}_B$, which we break into components as in Eq.~\eqref{eqn:bundle_mat}.
We now perform an algebraic decomposition of each of these pieces:

\begin{subequations} \label{eqn:alg_decomp}
  \begin{align}
    \smallunderset{PP}{A}^a{}_b &\equiv \smallunderset{[PP]}{A}^a{}_b + \smallunderset{\langle PP\rangle}{A}^a{}_b + \frac{1}{4} \smallunderset{PP}{A} \delta^a{}_b, \displaybreak[0] \\
    \smallunderset{PJ}{A}{}^a{}_{bc} &\equiv 2 \smallunderset{PJ}{A}{}_{[b} \delta^a{}_{c]} + \big({}^\dagger \smallunderset{\ddagger PJ}{A}\big){}^a{}_{bc} + \smallunderset{\langle PJ\rangle}{A}{}^a{}_{bc}, \displaybreak[0] \\
    \smallunderset{JP}{A}{}^{ab}{}_c &\equiv 2 \smallunderset{JP}{A}{}^{[a} \delta^{b]}{}_c + \big({}^\dagger \smallunderset{\ddagger JP}{A}\big){}^{ab}{}_c + \smallunderset{\langle JP\rangle}{A}{}^{ab}{}_c, \displaybreak[0] \\
    \smallunderset{JJ}{A}{}^{ab}{}_{cd} &\equiv 2 \delta^{[a}{}_{[c} \smallunderset{JJ} A^{b]}{}_{d]} + \smallunderset{[JJ]}{A}^{ab}{}_{cd} + \smallunderset{\langle JJ\rangle}{A}^{ab}{}_{cd} \nonumber \\
                                     &\hspace{1em}+ \smallunderset{*JJ}{A} \epsilon^{ab}{}_{cd}.
\end{align}
We also decompose $\smallunderset{JJ}{A}^a{}_b$ in the second-to-last line as

\begin{align}
    \smallunderset{JJ}{A}^a{}_b &\equiv \smallunderset{[JJ]}{A}^a{}_b + \smallunderset{\langle JJ\rangle}{A}^a{}_b + \frac{1}{4} \smallunderset{JJ}{A} \delta^a{}_b.
  \end{align}
\end{subequations}
These algebraically irreducible pieces have the following properties:

\begin{enumerate}

\item $\smallunderset{[PP]}{A}^a{}_b$ and $\smallunderset{[JJ]}{A}^a{}_b$ are antisymmetric, and have 6 independent components each;

\item $\smallunderset{\langle PP\rangle}{A}^a{}_b$ and $\smallunderset{\langle JJ\rangle}{A}^a{}_b$ are symmetric and trace-free, and have 9 independent components each;

\item $\smallunderset{\langle JP\rangle}{A}{}^{ab}{}_c$ and $\smallunderset{\langle PJ\rangle}{A}{}^a{}_{bc}$ are trace-free on all indices and satisfy

  \begin{equation}
    \smallunderset{\langle PJ\rangle}{A}{}_{[abc]} = \smallunderset{\langle JP\rangle}{A}{}_{[abc]} = 0,
  \end{equation}
  implying they have 16 independent components each;

\item $\smallunderset{[JJ]}{A}^{ab}{}_{cd}$ is trace-free on all indices and antisymmetric on interchange of the first two and last two indices, so it has 9 independent components; and

\item $\smallunderset{\langle JJ\rangle}{A}^{ab}{}_{cd}$ is trace-free on all indices, symmetric on interchange of the first two and last two indices, and satisfies

  \begin{equation}
    \smallunderset{\langle JJ\rangle}{A}{}_{[abcd]} = 0,
  \end{equation}
  giving it 10 independent components.

\end{enumerate}
The following results show how to construct the algebraically irreducible pieces from the full matrix $A^A{}_B$:

\begin{subequations} \label{eqn:alg_xx_def}
  \begin{align}
    \smallunderset{[xx]}{A}{}_{ab} &= \smallunderset{xx}{A}{}_{[ab]}, \\
    \smallunderset{xx}{A} &= \smallunderset{xx}{A}^a{}_a, \\
    \smallunderset{\langle xx\rangle}{A}{}_{ab} &= \smallunderset{xx}{A}{}_{(ab)} - \frac{1}{4} g_{ab} \smallunderset{xx}{A}, \\
    \smallunderset{JJ}{A}^a{}_b &= \smallunderset{JJ}{A}^{ac}{}_{bc} + \frac{1}{6} \smallunderset{JJ}{A}^{cd}{}_{cd} \delta^a{}_b,
  \end{align}
\end{subequations}
where $x$ is either $P$ or $J$,

\begin{subequations} \label{eqn:alg_xy_def}
  \begin{align}
    \smallunderset{PJ}{A}{}_a &= -\frac{1}{3} \smallunderset{PJ}{A}^b{}_{ba}, \\
    \smallunderset{JP}{A}{}^a &= -\frac{1}{3} \smallunderset{JP}{A}^{ab}{}_b, \\
    \smallunderset{\ddagger xy}{A}{}_a &= \big({}^\ddagger \smallunderset{xy}{A}\big){}_a, \\
    \smallunderset{\langle PJ\rangle}{A}^a{}_{bc} &= \smallunderset{PJ}{A}{}^a{}_{bc} - \smallunderset{PJ}{A}{}_{[b} \delta^a{}_{c]} + \epsilon^a{}_{bcd} \smallunderset{\ddagger PJ}{A}^d, \\
    \smallunderset{\langle JP\rangle}{A}{}^{ab}{}_c &= \smallunderset{JP}{A}{}^{ab}{}_c - \smallunderset{JP}{A}{}^{[a} \delta^{b]}{}_c + \epsilon^{ab}{}_{cd} \smallunderset{\ddagger JP}{A}^d,
  \end{align}
\end{subequations}
where $x\neq y$ is either $P$ or $J$, and

\begin{subequations} \label{eqn:alg_jj_def}
  \begin{align}
    \smallunderset{[JJ]}{A}^{ab}{}_{cd} &= \frac{1}{2} \left(\smallunderset{JJ}{A}^{ab}{}_{cd} - \smallunderset{JJ}{A}{}_{cd}{}^{ab}\right) - 2 \delta^{[a}{}_{[c} \smallunderset{[JJ]}{A}^{b]}{}_{d]}, \\
    \smallunderset{*JJ}{A} &= -\frac{1}{24} \epsilon^{abcd} \smallunderset{JJ}{A}{}_{abcd}, \\
    \smallunderset{\langle JJ\rangle}{A}^{ab}{}_{cd} &= \frac{1}{2} \left(\smallunderset{JJ}{A}^{ab}{}_{cd} + \smallunderset{JJ}{A}{}_{cd}{}^{ab}\right) - 2 \delta^{[a}{}_{[c} \smallunderset{\langle JJ\rangle}{A}^{b]}{}_{d]} \nonumber \\
                                        &\hspace{1em}- \frac{1}{2} \delta^{[a}{}_{[c} \delta^{b]}{}_{d]} \smallunderset{JJ}{A} - \smallunderset{*JJ}{A} \epsilon^{ab}{}_{cd}.
  \end{align}
\end{subequations}

There are two main uses of this decomposition.
The first is that many of these pieces have a physically relevant meaning.
For example, assuming that $J^{ab} = 0$, then $\smallunderset{[PP]}{A}^a{}_b$, $\smallunderset{\langle PP\rangle}{A}^a{}_b$, and $\smallunderset{\langle PP\rangle}{A}$ can be understood as an infinitesimal rotation, shear, and expansion of $P^a$, respectively (the latter two transformations change the rest mass $P^a P_a$).
As another example, $\smallunderset{JP}{A}^a$ is the term that contributes to the change in $J^{ab}$ in flat spacetime from a change of origin.

The second main use of this decomposition is that certain of these irreducible pieces may vanish for particular matrices; this could make it easier to compute the number of independent components that these matrices have.
For example, in the case where $A^A{}_B = \delta^A{}_B - \mathring{\Lambda}{}^A{}_B$, we can easily see from Eq.~\eqref{eqn:0_holonomy_mat} that the only nonzero pieces are $\smallunderset{[PP]}{A}^a{}_b = \smallunderset{[JJ]}{A}^a{}_b$ and $\smallunderset{JP}{A}^a$; thus, the holonomy has only 10 independent components.
Similarly, if we set $A^A{}_B = \vkappa{R}{}^A{}_{Bcd} \dot{\gamma}^c \dot{\bar \gamma}^d$ (an infinitesimal version of the holonomy for arbitrary $\varkappa$), we can easily show from Eq.~\eqref{eqn:mathfrak_kappa_R} and the symmetries of $\vkappa{K}_{abcd}$ in Eq.~\eqref{eqn:K_bianchi} that

\begin{subequations}
  \begin{align}
    \smallunderset{JP}{A}^{ab}{}_c &= 0, \\
    \smallunderset{PP}{A} &= \smallunderset{JJ}{A} = \smallunderset{* JJ}{A} = 0, \\
    \smallunderset{\ddagger PJ}{A}^a &= 0.
  \end{align}
\end{subequations}
This matrix then can have at most 69 independent components (it has fewer, but the algebraic decomposition only gives us an upper bound).
For the general case of the holonomy for arbitrary $\varkappa$ around a narrow loop, the algebraic decomposition gives no additional information about the number of independent components.

\bibliography{Refs}

\end{document}